\documentclass[10pt,aps,prd,a4paper,nofootinbib,
preprintnumbers,superscriptaddress,showpacs,showkeys,twocolumn]{revtex4-2}

\usepackage{amsmath}
\usepackage{amssymb}
\usepackage{bm}
\usepackage{graphicx}
\usepackage{xcolor}
\usepackage{slashed}
\usepackage{comment}
\usepackage[normalem]{ulem}

\usepackage[sort,compress]{cleveref}

\crefname{section}{Sec.\!}{Secs.\!}
\crefname{equation}{Eq.\!}{Eqs.\!}
\crefname{figure}{Fig.\!}{Figs.\!}
\crefname{table}{Tab.\!}{Tabs.\!}
\crefname{appendix}{App.\!}{Apps.\!}
\crefname{chapter}{Chapter}{Chapters}

\crefname{section}{Sec.\!}{Secs.\!}
\crefname{equation}{Eq.\!}{Eqs.\!}
\crefname{figure}{Fig.\!}{Figs.\!}
\crefname{table}{Tab.\!}{Tabs.\!}
\crefname{appendix}{App.\!}{Apps.\!}

\newcommand{\be}{\begin{equation}}
\newcommand{\ee}{\end{equation}}
\newcommand{\ba}{\begin{eqnarray}}
\newcommand{\ea}{\end{eqnarray}}

\begin{document}

\title{Entropy from decoherence: a case study using glasma-based occupation numbers }

\author{Gabriele Coci} \email{gabriele.coci@dfa.unict.it}
\affiliation{Department of Physics and Astronomy "Ettore Majorana", University of Catania, Via Santa Sofia 64, I-95123 Catania, Italy}\affiliation{INFN-LNS Laboratori Nazionali del Sud, Via S. Sofia 62, I-95123 Catania, Italy}

 \author{Gabriele Parisi} \email{gabriele.parisi@dfa.unict.it}
\affiliation{Department of Physics and Astronomy "Ettore Majorana", University of Catania, Via Santa Sofia 64, I-95123 Catania, Italy}\affiliation{INFN-LNS Laboratori Nazionali del Sud, Via S. Sofia 62, I-95123 Catania, Italy}

\author{Salvatore Plumari} \email{salvatore.plumari@dfa.unict.it}
\affiliation{Department of Physics and Astronomy "Ettore Majorana", University of Catania, Via Santa Sofia 64, I-95123 Catania, Italy}\affiliation{INFN-LNS Laboratori Nazionali del Sud, Via S. Sofia 62, I-95123 Catania, Italy}

\author{Marco Ruggieri}\email{marco.ruggieri@dfa.unict.it}
\affiliation{Department of Physics and Astronomy "Ettore Majorana", University of Catania, Via Santa Sofia 64, I-95123 Catania, Italy}\affiliation{INFN-Sezione di Catania, Via Santa Sofia 64, I-95123 Catania, Italy}

\begin{abstract}
We compute the entropy-per-particle, $S/N$, produced by the 
decoherence of a coherent state interacting with an environment, 
using an analytical open quantum system approach. 
The coherent state considered is characterized by occupation numbers borrowed from the glasma fields produced in the early stages of high-energy nuclear collisions. 
The environment is modeled as the vacuum, 
and decoherence arises from the interaction of the state with vacuum fluctuations. 
We describe the system-environment interaction 
via a phase-damping model, which represents continuous measurements on the system without altering its energy 
or particle number. Starting from the occupation numbers typical of the Glasma in high-energy proton-nucleus and nucleus-nucleus collisions, we find that the final $S/N$ after decoherence is lower than that of a two-dimensional thermal bath of ultrarelativistic gluons, except for proton-nucleus collisions at small values of $g\mu$. 
Our results indicate that quantum decoherence alone does not generate sufficient entropy to transform the initial coherent state into a thermalized gluon bath.
\end{abstract}



\maketitle
\section{Introduction}

The production of entropy, when associated with the loss of information during the dynamical evolution of a system characterized by a complex quantum state, is a hot topic in modern physics. Its study is relevant both in Cosmology - where it is believed that the Universe underwent a phase transition from a vacuum state to a ``thermalized'' state at the end of cosmic inflation following the Big Bang - and in Nuclear Physics, where
the so-called Little Bang is studied, referring to the formation of a strongly interacting state of matter known as Quark-Gluon Plasma in heavy-ion collisions~\cite{Geiger:1994ip,Reiter:1998uq,PhysRevLett.56.1663,Kunihiro:2008gv,Fries:2009wh,PhysRevC.76.024910,Ivanov:2016hes,Muller:2011ra}.

It is well known that 
quantum decoherence produces entropy~\cite{Zurek:1994wd,Elze:1994qa}. 
Within this framework,
a quantum system initially in a pure state 
is characterized by a density operator, $\rho$,
that is idempotent and satisfies the condition
$\mathrm{Tr}(\rho) = \mathrm{Tr}(\rho^2) =1$,
and has a vanishing von Neumann entropy, $S$, 
defined as
\begin{equation}
S = -\mathrm{Tr}(\rho \log\rho ).
\label{eq:intenseplex_intro}
\end{equation}
When the system is coupled to 
an external environment (typically modeled as a thermal
bath, but it can also be an ensemble of quantum fluctuations
and vacuum fluctuations), it evolves into an incoherent mixture,
in which $\mathrm{Tr}(\rho^2) <1$ and $S>0$.
This is due to the randomization of the relative phases
of the states that form the mixture, as a result of the
interactions with the environment, which implies
the decay of the off-diagonal elements of $\rho$.
This process, that eventually leads to an incoherent mixture
of states, is called quantum decoherence.
The theory of open quantum systems, which has been recently applied in the context of heavy-ion physics~\cite{Rais:2025fps,Delorme:2024rdo,Neidig:2023kid,Brambilla:2020qwo,Kajimoto:2017rel,Brambilla:2016wgg,Akamatsu:2011se,Blaizot:2017ypk,Katz:2015qja,DeJong:2020riy}
provides the ground basis to study decoherence processes through quantum master equation, which describes the evolution of a system interacting with a large environment.
The timescale over which decoherence takes place depends on the coupling of the system to the environment, as well as on the microscopic properties of the environment itself.
The evolution of the system density operator $\rho$ can be studied, at least in principle, by solving a quantum master equation. This can be derived from the Liouville-von Neumann equation of the total density operator by tracing over the degrees of freedom of the environment. The resulting equation is challenging to solve and, more critically, it is not trace preserving, thereby it violates the requirement of probability conservation $\mathrm{Tr}(\rho)=1$. Under the commonly adopted Born–Markov approximation, the system dynamics reduces to a dissipative master equation: in the Markovian limit, the evolution equation for $\rho(t)$ assumes the form of the well-known Gorini–Kossakowski–Sudarshan–Lindblad (GKSL), or simply Lindblad equation.~\cite{Lindblad:1975ef,Gorini:1975nb}. 
This type of derivation is matter of textbooks, see for example~\cite{Carmichael::385006,Breuer:2007juk}.
The solution of the Lindblad equation allows one to estimate the timescale for the decoherence, as well as to compute the entropy produced asymptotically by this process.

The simplest frameworks to study decoherence are 
phase-damping and amplitude-damping models.
In the former, the 
interaction of the system with the environment does not
change the
occupation numbers. Hence, within these models there is
no energy and particle exchange between the system and
the environment. 
For example, when this model is applied to the decoherence of a coherent state of the harmonic oscillator, the value of the von Neumann entropy
at equilibrium depends only on the occupation number 
of the coherent state and not on the temperature, or other properties, of the environment~\cite{Vidiella-Barranco:2016hnh,Estes:1968quantum}. 
On the other hand,
in amplitude-damping models, the decay of the off-diagonal
elements of the density operator is accompanied by the 
exchange of particles and energy between the system and
the reservoir. Consequently, the occupation numbers
evolve with time during the 
decoherence. In this case, the average value of
the particle number of the system equilibrates to that
of the environment. If the environment is in thermal
equilibrium, then amplitude-damping models lead at
thermalization, for detailed treatment see Refs.~\cite{Carmichael::385006,Davies:1976quantum,Mandel:1995optical}.

The purpose of the study presented in this article is the
computation of the decoherence entropy of a coherent state within
the phase-damping model.
In particular, the occupation number of the coherent state is borrowed from the Glasma picture of high-energy nuclear collisions~\cite{McLerran:1993ni,McLerran:1993ka,McLerran:1994vd,Iancu:2000hn}, for reviews see Refs.~\cite{Fukushima:2011nq,Gelis:2010nm,Iancu:2003xm,McLerran:2008es,Gelis:2012ri}.
Within this picture, the early stage of the system produced by the
collision is made of many gluons forming a coherent state.
This state then evolves through non-abelian dynamics of the
Yang-Mills theory.
We define the occupation numbers from the
leading-order glasma fields, from which we define a coherent state
in momentum space. Then, we use the phase-damping model
to compute the amount of entropy produced by the quantum
decoherence of this state.
For the sake of nomenclature, we refer to the coherent state as the Glasma, and to its real-time evolution as the evolving Glasma. This despite the fact that we do not attempt to really evolve the system by solving the Yang-Mills equations, as recently applied in other studies in the context of heavy-ion collisions~\cite{Avramescu:2024xts,Oliva:2024rex,Pooja:2023gqt,Avramescu:2023qvv,Sun:2019fud,Liu:2019lac,Boguslavski:2020tqz,Carrington:2020ssh,Carrington:2021qvi,Iida:2014wea,Matsuda:2022hok,Tsukiji:2017pjx,Tsukiji:2016krj,Iida:2013qwa}.

Within our study, we assume that the process of decoherence takes place thanks to the continuous interaction of the  system with the quantum vacuum fluctuations of the environment.
We then compare the ratio $S/N$, where $S$ is the entropy produced by decoherence and $N$ is the total occupation number of the coherent state, with that of a thermalized
gluon gas. We take the difference between the two ratios as a measure of the amount of thermalization produced by decoherence.

In this work we do not study the transient, as it requires
the calculation of the decoherence time, $\tau_\mathrm{d}=1/\gamma$,
which would be highly 
model-dependent. 
While model-building is certainly interesting, 
we prefer to focus on the total amount of entropy
produced by the process of quantum decoherence for this
particular coherent state. 
In particular, decoherence can occur because of the continuous probing of the coherent glasma fields by the environment, in the simplest case this being vacuum fluctuations.
Typical vacuum fluctuations in the QCD vacuum should happen on a timescale
$O(1/\Lambda_\mathrm{QCD})$; moreover, in the glasma fields a natural
energy scale is present, which is the saturation scale $Q_s$ to which the
natural glasma timescale, $O(1/Q_s)$, is related.
Hence,
in the context of the Glasma that we analyze here,
it is reasonable to assume $\tau_\mathrm{d}$ to be in the range $(1/Q_s,1/\Lambda_\mathrm{QCD})$.
An estimate of the decoherence time of the Glasma, coming from the non-abelian interactions rather than from a pure decoherence due to continuous interactions with the vacuum fluctuations, 
has been given in~\cite{Muller:2005yu,Muller:2011ra}, 
where it has been found $\tau_d \sim Q_s^{-1}$. 
This is a natural result as $Q_s$ is the only energy scale in the problem. In addition to the quantum decoherence itself, it is worth mentioning that
the non-abelian interaction of the Glasma with fluctuations arising at order
$\alpha_s$ generates further entropy~\cite{Iida:2014wea,Tsukiji:2016krj,Tsukiji:2017pjx,Matsuda:2022hok}.

This article is organized as follows: in Sec.~II we describe the open quantum master equation, which allows for an analytical derivation of quantum decoherence from the dynamical evolution of a coherent state. In Sec.~III we show how we can map the initial state of a boost-invariant field configuration, the Glasma, onto a coherent state. 
In Sec.~IV we compute the occupation numbers of the 
coherent state. 
The entropy we are interested in is the von Neumann entropy: whenever we use the term entropy we refer to that, unless differently stated. 
Finally, we present our results about decoherence entropy in the asymptotic limit and give our conclusions.
In this work we use $\hbar=c=k_B=1$.

\section{The decoherence of a coherent state within phase-damping models\label{sec:pcm}}
In this section, we briefly review the quantum decoherence of a single coherent state (CS) of a simple harmonic oscillator (SHO), within the phase-damping model. In this model, decoherence takes place thanks to the interaction of the system, namely the CS, with an external environment, that we call the reservoir. 
The full hamiltonian is written as $H = H_S + H_R + H_{SR}$, where
\begin{eqnarray}
&& H_S = \omega_0 a^\dagger a,\label{eq:ho_ff_1}\\
&& H_R = \sum_j \omega_j b_j^\dagger b_j,\label{eq:ho_ff_2}\\
&&H_{SR} = 
a^\dagger a
\sum_j (\kappa^*_j b^{\dagger}_j + \kappa_j b_j) = 
a^\dagger a
(\Gamma^\dagger +\Gamma).\label{eq:ho_ff_3}
\end{eqnarray}
Here, $S$ and $R$ denote the system and the reservoir, respectively.
$\omega_0$ denotes the proper frequency of the system, $a^\dagger$ and $a$ are the ladder operators of the HO. In this model, $R$ is a collection of 
harmonic oscillators with characteristic frequencies $\omega_j$ 
and creation and annihilation operators $b^\dagger_j$ and $b_j$. 
Finally, $H_{SR}$ describes the coupling of $S$ to $R$.
In particular, $S$ is coupled to each harmonic oscillator $j$ of $R$ through the coupling constant $\kappa_j$.
In this model, the number operator $N_a=a^\dagger a$ of the oscillator $a$ commutes with the hamiltonian, therefore $\langle N_a \rangle$ is unchanged in the evolution.

\subsection{Quantum decoherence of a coherent state}

To begin with, we review the idea of the CS of a SHO.
To this end, let us consider an oscillator with
characteristic frequency $\omega$, with hamiltonian
\begin{equation}
H = \omega_0 a^\dagger a,
\label{eq:musi1}
\end{equation}
where $a^\dagger$, $a$ correspond to the standard ladder operators,
that satisfy the relations
\begin{equation}
a^\dagger |n\rangle = \sqrt{n+1}|n+1\rangle,~~~
a |n\rangle = \sqrt{n}|n-1\rangle,
\label{eq:laportaeraaperta}
\end{equation}
as well as the canonical commutation relation
\begin{equation}
[a,a^\dagger]=1.
\label{eq:astra}
\end{equation}
Note that in Eq.~\eqref{eq:musi1} we subtracted the zero mode energy,
which is irrelevant in our problem.

The CS of the harmonic oscillator Eq.~\eqref{eq:musi1}
is defined as the eigenstate of the annihilation operator~\cite{Glauber:1963fi}, namely
\begin{equation}
a|\alpha\rangle = \alpha |\alpha\rangle;
\label{eq:st_coe_3}
\end{equation}
here $\alpha$ is, in general, a complex number,
and 
\begin{equation}
\langle \alpha | a |\alpha\rangle = \alpha.
\label{eq:alpha_exp_val_1107}
\end{equation}
The CS can be expressed in terms of the basis of the Fock space as
\begin{equation}
|\alpha\rangle = e^{-|\alpha|^2/2} \sum_n \frac{\alpha^n}{\sqrt{n!}} |n\rangle.
\label{eq:expa_fock}
\end{equation}
One way to read Eqs.~\eqref{eq:st_coe_3} 
and~\eqref{eq:expa_fock} is that 
removing one quantum from the CS, by formally applying $a$ to the state,
results in the same state apart from the change of the norm. 
The average occupation number in the CS~\eqref{eq:expa_fock} is 
\begin{equation}
n \equiv \langle\alpha|a^\dagger a|\alpha\rangle=|\alpha|^2.
\label{eq:nmedio}
\end{equation}

In the Fock space, the density matrix operator for the CS~\eqref{eq:expa_fock} is given by
\begin{equation}
\rho = \sum_{m,n}\rho_{mn}|m\rangle\langle n |,
\end{equation}
with elements
\begin{equation}
\rho_{mn} = e^{-|\alpha|^2}
\frac{\alpha^m(\alpha^*)^n}{\sqrt{n!m!}}=
e^{-|\alpha|^2}
\frac{\langle a\rangle^m\langle a^\dagger \rangle^n}{\sqrt{n!m!}}.
\label{eq:centramaterazzi}
\end{equation}

In the phase-coupling model, the 
quantum master equation for the density operator $\rho$ in the Schrodinger picture is
easily obtained following the standard techniques of open quantum systems,
tracing over the environment degrees of freedom and assuming
Born-Markov approximation, see for example~\cite{Breuer:2007juk,Carmichael::385006}.
This procedure leads to the following Lindblad equation,
\begin{eqnarray}
\dot\rho &=& -i\omega^\prime_0[a^\dagger a,\rho]\nonumber\\
&&+
\frac{\gamma}{2}(1+\bar n)(2a^\dagger a\rho a^\dagger a - 
a^\dagger a a^\dagger a\rho - \rho a^\dagger a a^\dagger a)
\nonumber \\
&&+\frac{\gamma}{2}\bar n(2a^\dagger a \rho a^\dagger a  
- a^\dagger a a^\dagger a\rho - \rho a^\dagger aa^\dagger a),
\label{eq:lindbGHJKL}
\end{eqnarray}
where
\begin{eqnarray}
\bar n &=&\bar n(\omega_0),\label{eq:forN}\\
\gamma &=&2\pi g(\omega_0)|\kappa(\omega_0)|^2.\label{eq:forGAMMA}
\end{eqnarray}
In particular, the functions $g(\omega)$ and $\kappa(\omega)$ characterize the time correlations of the reservoir operators,
\begin{eqnarray}
\langle \Gamma^\dagger(t) \Gamma(t-\tau) \rangle_R &=&
\int_0^\infty d\omega~e^{i\omega\tau}~g(\omega)|\kappa(\omega)|^2
\bar n(\omega),\label{eq:correlators_aaa_1}\\ 
\langle \Gamma(t) \Gamma^\dagger(t-\tau) \rangle_R&=&
\int_0^\infty d\omega~e^{i\omega\tau}~g(\omega)|\kappa(\omega)|^2
\left[1+\bar n(\omega)\right],
\nonumber\\
\label{eq:correlators_aaa}
\end{eqnarray} 
where $\langle\rangle_R$ denotes the ensemble average over the reservoir.
In Eqs.~\eqref{eq:forN}-\eqref{eq:correlators_aaa} 
we changed the summation over the collection of h.o. 
with frequencies $\omega_j$ into an integration, 
introducing a density function of states $g(\omega) d\omega$ which gives the number of h.o. having frequencies in the interval $\omega$ to $\omega + d\omega$. 
As commonly done in the literature, 
for the sake of simplicity
we work under the assumption that
$R$ is a Markovian reservoir, so the correlators Eqs.~\eqref{eq:correlators_aaa_1}~\eqref{eq:correlators_aaa}
are proportional to the $\delta(\tau)$: 
this can be obtained by assuming a specific form 
for $g(\omega)|\kappa(\omega)|^2$,
see for example \cite{Carmichael::385006}.
We do not aim at justifying
this hypothesis here; rather, we take it as a way to simplify the formulation
of the problem. Extension to the case of a non-Markovian reservoir
is far from being trivial; in fact, how to do this extension properly
is still a matter of debate in the literature.

The term $\omega_0^\prime$ on the right hand side
of Eq.~\eqref{eq:lindbGHJKL} takes into account a potential
renormalization of $\omega_0$ induced by the interaction with the
reservoir. The shift $\omega_0^\prime - \omega_0$
is commonly referred to as the Lamb shift~\cite{Carmichael::385006}.
In our study, this shift is not important, because eventually
we want to couple the CS to the vacuum, and in this case it is well known
that the correlator responsible for the shift vanishes.
Hence, we
neglect it from now on.

We notice that the terms proportional to $\gamma$
in Eq.~\eqref{eq:lindbGHJKL} lead to a non-unitary evolution
of the density operator; that is, the evolution of $\rho$
is
characterized by diffusion and dissipation.
Within the context of our study, the dissipation corresponds to 
the degradation of the quantum coherence of the initial state.
Interestingly, the non-unitary  evolution 
happens also when the CS is coupled
to the vacuum: in this case, the decoherence is induced
by the interaction of the system with the quantum fluctuations. 
For this case, $\bar n(\omega) =0$ and 
only the correlations Eq.~\eqref{eq:correlators_aaa} 
contribute. If the reservoir is made of vacuum fluctuations only,
we can therefore simplify the master equation as
\begin{eqnarray}
\dot\rho &=& -i\omega^\prime_0[a^\dagger a,\rho]\nonumber\\
&&+
\frac{\gamma}{2} (2a^\dagger a\rho a^\dagger a - 
a^\dagger a a^\dagger a\rho - \rho a^\dagger a a^\dagger a).
\label{eq:lindbGHJKLvacuum}
\end{eqnarray}

The formal solution of Eq.~\eqref{eq:lindbGHJKLvacuum}
with initial condition given by the CS Eq.~\eqref{eq:expa_fock}
is easily expressed in the Fock basis as~\cite{Walls:1985tm}
\begin{equation}
\rho(t)=\sum_{m,n}\rho_{mn}(t)|n\rangle\langle m |,
\label{eq:monzamanchester}
\end{equation}
where
\begin{equation}
\rho_{mn}(t) = e^{-\gamma(n-m)^2t/2}\rho_{mn}(0),
\label{eq:acasaunpunto}
\end{equation}
and $\rho_{mn}(0)$ is given by Eq.~\eqref{eq:centramaterazzi}.
We notice that the off-diagonal terms of $\rho(t)$ 
in Eq.~\eqref{eq:acasaunpunto}
exponentially
decay with characteristic time
\begin{equation}
\tau^{mn}_d = \frac{2}{\gamma(m-n)^2},~~~m\neq n.
\label{eq:tau_d_mn_uuu}
\end{equation}
This is precisely the quantum decoherence phenonemon arising from the 
coupling of the
CS to the reservoir.
As a consequence, at asymptotically large times the density
operator becomes
\begin{equation}
\rho_\infty = e^{-|\alpha|^2}\sum_n
\frac{|\alpha|^{2n}}{n!}|n\rangle\langle n |,
\label{eq:quantopesanoitrepuntididomanisera}
\end{equation}
namely, an incoherent mixture of eigenstates of the number operator
with weights equal to those of the
coherent state, see Eq.~\eqref{eq:centramaterazzi}.

The time dependence of the average of the ladder operator in this model is easily found to be 
\begin{equation}
\langle a \rangle = \mathrm{Tr}(a\rho(t)) = \alpha e^{-\gamma t/2}.
\label{eq:comeabbiamosemprefatto}
\end{equation}
Moreover,
\begin{equation}
\langle a^\dagger a \rangle = \mathrm{Tr}(a^\dagger a \rho(t)) = 
|\alpha|^2,
\label{eq:perquantoriguardagliattaccanti}
\end{equation}
namely, the expectation value of the number operator is unaffected by the coupling with the reservoir.
In this model, the interaction with the reservoir causes the loss of coherence without the exchange of particles and energy between the system and the reservoir.

We notice that the result Eq.~\eqref{eq:quantopesanoitrepuntididomanisera} is in agreement with the assumption used in~\cite{Iida:2014wea},
where the Glasma has been mapped to a decoherent ensemble of gluons
with each weight given at each time by that of the coherent state.

The evolution of the density operator in the presence of decoherence additionally leads to the production of entropy. In this context, entropy is identified with the von Neumann entropy,
\begin{equation}
S = -\mathrm{Tr}[\rho(t)\log\rho(t)].
\label{eq:intenseplex}
\end{equation}
Within this model, the occupation number of the CS is not affected by the reservoir. 
Moreover, the coupling to the reservoir enters only via $\gamma$, that sets the timescale for the decoherence.
Consequently, the asymptotic value of $S$ does not depend on the environment, rather on the $|\alpha|^2$ of the CS.

This can be explicitly proved by computing $S$ in
the $\gamma t \rightarrow\infty$ limit.
In fact, in this limit the off-diagonal elements of the density operator vanish, and the $\ell^{th}$ diagonal element corresponds to the $\ell^{th}$ eigenvalue of the density operator, which is
\begin{equation}
\lambda_{\ell} = e^{-n}\frac{n^\ell}
{\ell!},~~~ n = |\alpha|^2.
\label{eq:rigoresudumfriesNETTO}
\end{equation}
This implies that the decoherence entropy, $S_\infty$, of the initial CS is
\begin{equation}
S_\infty = -\sum_\ell e^{-n}\frac{n^\ell}
{\ell!}\log \left(e^{-n}\frac{n^\ell}
{\ell!}\right).\label{eq:nonmimeraviglia}
\end{equation}
Similarly, the entropy per particle produced by decoherence is
\begin{equation}
\frac{S_\infty}{n} = 
1-\log n  + \frac{e^{-n }}{n }\sum_{\ell=1}^\infty
\frac{n^\ell}{\ell !}\log\ell!.
\label{eq:entrperparteq}
\end{equation}

\section{The coherent state from the Glasma}

In this section, we define and compute the occupation numbers that we will use to model the Glasma as a coherent state.

\subsection{Statement of the problem}
Our idea is to build up a CS at $t=0$ from the occupation numbers of the Glasma, then compute the decoherence entropy of this state within the phase-coupling model described in the previous section. 
Within our approach, the environment constantly interacting with the coherent state is the vacuum, 
that probes the system via quantum fluctuations.
In our formulation, we follow Ref.~\cite{Iida:2014wea}, assuming normal-mode-like relations between the quantum field operators of the Glasma and the ladder operators. Then, we relate the expectation values of these operators to the occupation numbers of the coherent state.

To begin with, we write the field operators for the color fields in a quantization volume $V = L_T^2  \times L_z$ as
\begin{eqnarray}
A^{ai}(x) &=&\frac{1}{V}\sum_{\bm k} \frac{1}{\sqrt{2\omega_k}}
\left\{
e^{i \bm k \cdot \bm x }  a_{kai}   +
e^{-i \bm k \cdot \bm x  }  a^\dagger_{kai}
\right\}, \label{eq:ivanolombardo1}\\
E^{ai}(x) &=&\frac{1}{V}\sum_{\bm k} \frac{-i\omega_k}{\sqrt{2\omega_k}}
\left\{
e^{i \bm k \cdot \bm x }  a_{kai}   -
e^{-i \bm k \cdot \bm x  }  a^\dagger_{kai}
\right\}. \label{eq:ivanolombardo2}
\end{eqnarray}
In high-energy nuclear collisions, the $z$-direction is that of the flight of the
two colliding objects, that we call the longitudinal direction. The plane perpendicular to $z$ is dubbed the transverse plane. Within our scheme,
the quantization volume has extension $L_z$ along the
longitudinal direction and $L_T^2$ in the transverse plane.
In Eqs.~\eqref{eq:ivanolombardo1}~\eqref{eq:ivanolombardo2} $\bm k = (k_x,k_y,k_z)$ 
denotes momentum, $a=1,\dots,N_c^2-1$ is the adjoint color index labelling the gluon fields.
Finally, $a_{kai}$ and $a^\dagger_{kai}$ denote
the ladder operators for the mode $kai$ of the 
quantum field, with 
\begin{equation}
kai\equiv (k_x,k_y,k_z,a,i),
\label{eq:condensed?indices}
\end{equation}
and the characteristic frequency is
\begin{equation}
\omega_k = \sqrt{k_x^2 + k_y^2 + k_z^2}.
\label{eq:omega_kappa}
\end{equation}
From Eq.~\eqref{eq:ivanolombardo1} and~\eqref{eq:ivanolombardo2} we get
\begin{eqnarray}
a_{kai}  = \frac{1}{\sqrt{2\omega_k}}
\left(
\omega_k  \tilde A_{kai}   + 
i   \tilde E_{kai}  
\right),\label{eq:lllPPP}
\end{eqnarray}
where the Fourier transforms of the fields are
\begin{align}
\tilde A_{kai} &= \int d^2 \bm x_T dz~e^{-i\bm k_T \cdot \bm x_T}
e^{-i k_z z}
A^{ai}(\bm x_T,z),\label{eq:tra_A_1354_3}\\
\tilde E_{kai} &= \int d^2 \bm x_T dz~e^{-i\bm k_T \cdot \bm x_T}
e^{-i k_z z}
E^{ai}(\bm x_T,z).\label{eq:tra_E_1354_3}
\end{align}
Assuming a $z-$independent configuration of the
gauge fields, that mimicks the boost-invariant Glasma, we can limit ourselves to consider only
the $k_z=0$ modes in Eq.~\eqref{eq:tra_A_1354_3} and Eq.~\eqref{eq:tra_E_1354_3}.
From these, and by introducing the two-dimensional
Fourier transforms
\begin{align}
A_{kai} &= \int d^2 \bm x_T~e^{-i\bm k_T \cdot \bm x_T}
A^{ai}(\bm x_T),\label{eq:tra_A_1354}\\
E_{kai} &= \int d^2 \bm x_T~e^{-i\bm k_T \cdot \bm x_T}
E^{ai}(\bm x_T),\label{eq:tra_E_1354}
\end{align}
we can rewrite  Eq.~\eqref{eq:lllPPP} in the form
\begin{eqnarray}
a_{kai}  = \frac{L_z}{\sqrt{2\omega_k}}
\left(
\omega_k  A_{kai}   + 
i   E_{kai}  
\right),\label{eq:lllPPP_2D}
\end{eqnarray}
where $L_z$ comes from the trivial integration 
over the $z-$direction. Notice that in natural units, $A_{kai}$ and $E_{kai}$ carry dimensions of 
energy$^{-1}$ and energy$^{0}$ respectively, 
$a_{kai}$ carries dimensions of energy$^{-3/2}$. This is important to define correctly the occupation numbers of the CS from the Glasma,
see Eq.~\eqref{eq:grandpa_shark}.

For the standard glasma initialization, 
$A^{az}=0$, therefore we are left with two independent components of the gauge potential only. 
Moreover, $E^{ax}=E^{ay}=0$, so that the initial color-electric field is purely longitudinal.

In a CS corresponding to the state $kai$, 
the quantum expectation value of $a_{kai}$
is nonzero, see Eq.~\eqref{eq:alpha_exp_val_1107}.
Such coherent states can be written as
\begin{equation}
|\beta_{kai}\rangle 
=
e^{-|\beta_{kai}|^2/2}\sum_{n=0}^\infty
\frac{\beta_{kai}^n}{\sqrt{n!}}|n\rangle,
\label{eq:walkingcrab}
\end{equation}
where, from Eq.~\eqref{eq:lllPPP_2D}, 
\begin{align}
\beta_{kai} &= \langle \beta_{kai} |a_{kai}|\beta_{kai} \rangle,
\\
&=\frac{L_z}{\sqrt{2\omega_k}}
\left(
\omega_k  \langle \beta_{kai} |A_{kai} |\beta_{kai} \rangle  + 
i  \langle \beta_{kai} | E_{kai} |\beta_{kai} \rangle 
\right),\nonumber\\
&\label{eq:probably_1146}
\end{align}
and $|n\rangle$ denotes the Fock state with $n$ gluons in the quantum state $kai$.
The occupation number of the CS with quantum
numbers $kai$ is 
\begin{equation}
n_{kai} =\langle a^\dagger_{kai}a_{kai}\rangle.
\label{eq:occ_numb_1059}
\end{equation}
The density operator of the ensemble of
coherent states at $t=0$  is thus
\begin{equation}
\rho(t=0) = \prod_{kai}  |\beta_{kai}\rangle\langle\beta_{kai}|.
\label{eq:saraaaaaaaaaBBBBBBBaaaHHH}
\end{equation}

In order to map the glasma fields with the
ensemble of coherent states Eq.~\eqref{eq:saraaaaaaaaaBBBBBBBaaaHHH}
we follow Ref.~\cite{Iida:2014wea} and assume
that 
we can
take the classical limit of Eq.~\eqref{eq:lllPPP_2D},
replacing the quantum operators with their classical counterparts, namely with their expectation values on the coherent state.
Then, we calculate ensemble averages over color charges configurations that generate the 
Glasma in order to compute observables.
Within this assumption,
$a_{kai}$ for the glasma fields
is still given by Eq.~\eqref{eq:lllPPP_2D},
while $A_{kai}$ and $E_{kai}$ denote the
two-dimensional Fourier transforms of the
classical color fields of the Glasma,
that fluctuate on an event-by-event basis.

In practical calculations, it is much easier to
directly compute the ensemble-averaged
occupation number of the $kai$-CS,
\begin{equation}
n_{kai}=\langle|a_{kai}|^2 \rangle d^3k,
\label{eq:grandpa_shark}
\end{equation}
where $d^3k=dk_z d^2k_T$ and the brackets denote
the ensemble average.
The multiplication by $d^3k$ is introduced to 
make $n_{kai}$ a pure number.
Then, each CS from the Glasma can be identified by the characteristic value
\begin{equation}
\beta_{kai} = \sqrt{n_{kai}} e^{i\theta_{kai}}.
\label{eq:pic_pot_0_1}
\end{equation}
Here, $\theta_{kai}$ is an arbitrary phase that in principle
is different for each CS. For simplicity, we put $\theta_{kai}=0$,
as the decoherence entropy Eq.~\eqref{eq:nonmimeraviglia},
to which we are interested in our study,
depends only on the occupation number, hence on
$\langle|a_{kai}|^2\rangle$, and is insensitive to $\theta_{kai}$.

In the following section, we compute $a_{kai}$ and
$n_{kai}$ for the glasma fields. Then, we will use them in Section~\ref{sec:entro_prod_a} to compute the decoherence entropy within the phase-damping model.

\section{Occupation numbers of the coherent state}

In this section we compute $n_{kai}$ for the Glasma.
We split this section into two main parts. 
In Subsection~\ref{subsec:1} we compute $a_{kai}$,
see Eq.~\eqref{eq:lllPPP_2D}.
We then use this result in Subsection~\ref{subsec:2} to 
compute $n_{kai}$ according to Eq.~\eqref{eq:grandpa_shark}.
This whole section is quite technical, therefore the reader not interested in the details of the calculation can skip it entirely, as the main results will be summarized at the beginning of Section~\ref{sec:entro_prod_a}.
The calculations are based on the standard 
procedure used to construct the glasma fields in high-energy nuclear collisions.

\subsection{Calculation of $a_{kai}$\label{subsec:1}}

In order to prepare the initial coherent state from the Glasma produced in the high-energy collision of two projectiles, $A$ and $B$,
we firstly have to solve the Poisson equations
\begin{equation}
\nabla_\perp^2\Lambda_A = - \rho_A(\bm x_T),~~~
\nabla_\perp^2\Lambda_B = - \rho_B(\bm x_T),
\label{eq:GrazieFabian}
\end{equation}
where $\rho_A$ and $\rho_B$ 
denote the color charges of the two colliding objects.
In the above equations $\rho_{A,B}=\rho_{A,B}^a T_a$ and
 $\Lambda_{A,B}=\Lambda_{A,B}^a T_a$, where $T_a$ denote the
$SU(3)$-color generators.
These color charges are assumed to be gaussian random variables with zero average. Their correlators are in the form 
\begin{eqnarray}
&&\langle \rho_a(\bm x_T)   \rho_b(\bm y_T)  \rangle =
\langle\left[  g\mu \left(\bm v\right) \right]^2\rangle\delta_{ab} F(\bm u).
\label{eq:rai2HD_00:26}
\end{eqnarray}
Here, 
 $\bm v \equiv (\bm x_T + \bm y_T)/2$, and 
$\bm u = \bm x_T - \bm y_T$.
$\mu$ is an inverse length scale such that
$\mu^2$ measures the number of charge carriers per
unit of the transverse plane.
In modern implementations of the MV model, $\mu$
is coordinate-dependent\footnote{There is also a dependence
on spacetime rapidity, $\eta$. Here we limit ourselves to analyze the fields produced at midrapidity, $\eta=0$, therefore we do not consider any explicit $\eta-$dependence of $\mu$.}. Moreover, $g$ is the QCD coupling constant.
In most of the numerical studies, $F(\bm u)\propto \delta^2(\bm u)$, meaning that fluctuations of the color charges
are uncorrelated point by point in the transverse
plane.

The formal solution of Eqs.~\eqref{eq:GrazieFabian} can be written as
\begin{equation}
\Lambda_{A,B}(\bm x_T) = \int\frac{d^2q_T}{(2\pi)^2}
e^{i\bm q_T \cdot \bm x_T}\frac{\tilde\rho_{A,B}(\bm q_T)}
{q_T^2 + m^2}.
\label{eq:reggerbotta}
\end{equation}
Here $m$ is an infrared regulator. Formally, it is needed to regularize the inverse laplacian in the transverse coordinates. Physically, it can be understood as an effective way to remove the unphysical contributions of the colored fluctuations to the gauge potential on length scales larger than the nucleon size.
Typically $m$ is in the range $(0.1,0.4)$ GeV.

The gauge potential immediately after the collision is
\begin{equation}
A^i = i V\partial_i V^\dagger +  i W\partial_i W^\dagger,~~~i=x,y,
\label{eq:beneomale}
\end{equation}
where
\begin{equation}
V = e^{-i\Lambda_A},~~~W=e^{-i\Lambda_B}.
\label{eq:creareoccasioniinsiemeallasquadra}
\end{equation}
Taking into account the formal solution~\eqref{eq:reggerbotta},
it is easy to prove that the initial gauge potential~\eqref{eq:beneomale} can be written as
\begin{equation}
A^i(\bm x_T) = -\int \frac{d^2q_T}{(2\pi)^2}
e^{i\bm q_T \cdot \bm x_T}
\frac{iq_i [\tilde\rho_{A}(\bm q_T) + \tilde\rho_{B}(\bm q_T)]}
{q_T^2 + m^2},
\label{eq:igiocatorisonolenti}
\end{equation}
with $i=x,y$.
From this we immediately read
\begin{equation}
A^i(\bm q_T) = 
-\frac{iq_i [\tilde\rho_{A}(\bm q_T) + \tilde\rho_{B}(\bm q_T)]}
{q_T^2 + m^2},~~~i=x,y,
\label{eq:idatilihovisti}
\end{equation}
that is
\begin{equation}
A^{i}_a(\bm q_T) = 
-\frac{iq_i [\tilde\rho_{Aa}(\bm q_T) + \tilde\rho_{Ba}(\bm q_T)]}
{q_T^2 + m^2},~~~i=x,y.
\label{eq:lanostrasplendidachat}
\end{equation}

\begin{widetext}

Moreover, the initial longitudinal electric field is
\begin{equation}
E^z = i\left[
iV\partial_x V^\dagger,   iW\partial_x W^\dagger
\right] + i\left[
iV\partial_y V^\dagger,  i W\partial_y W^\dagger
\right]. 
\label{eq:devegiocareognivolta}
\end{equation}
Using $[T_a,T_b]=i f_{abc}T_c$, and following the same steps
that lead us to Eq.~\eqref{eq:idatilihovisti}, we get
\begin{equation}
E^z(\bm x_T) = \int \frac{d^2q_T}{(2\pi)^2}\int \frac{d^2k_T}{(2\pi)^2}
e^{i(\bm q_T + \bm k_T) \cdot \bm x_T}
\frac{(iq_x i k_x + iq_y i k_y)
[\tilde\rho_{Aa}(\bm k_T)   \tilde\rho_{Bb}(\bm q_T)]}
{(q_T^2 + m^2)(k_T^2 + m^2)}if_{abc}T_c.
\label{eq:ipoteridimarotta}
\end{equation}
From~\eqref{eq:ipoteridimarotta},
after reshuffling the indices to uniform the notation
with that 
of Eq.~\eqref{eq:lanostrasplendidachat},
we get the Fourier transform
\begin{equation}
E^z_a(\bm q_T) = -i\int\frac{d^2k_T}{(2\pi)^2}
\frac{ \bm k_T\cdot (\bm q_T-\bm k_T) 
[\tilde\rho_{Ab}(\bm k_T)   \tilde\rho_{Bc}(\bm q_T - \bm k_T)]}
{(k_T^2 + m^2)((\bm k_T-\bm q_T)^2 + m^2)} f_{bca}.
\label{eq:isuperpoteridimarotta}
\end{equation}
Then, the results~\eqref{eq:lanostrasplendidachat} 
and~\eqref{eq:isuperpoteridimarotta} can be used in Eq.~\eqref{eq:lllPPP_2D}
to produce the $a_{kai}$ of the initial coherent state. 
Taking into account that $A^i=0$ for $i=z$ and that
$E^i=0$ for $i=x,y$,
we get
\begin{eqnarray}
a_{kai} &=&
\frac{L_z}{\sqrt{2\omega_k}}
\left\{-\frac{i\omega_k k_i [\tilde\rho_{Aa}(\bm k_T) + \tilde\rho_{Ba}(\bm k_T)]}
{k_T^2 + m^2}(\delta_{ix}+\delta_{iy})\right.\nonumber\\
&&\left.
+\delta_{iz}\int\frac{d^2q_T}{(2\pi)^2}
\frac{ \bm q_T\cdot (\bm k_T-\bm q_T) 
[\tilde\rho_{Ab}(\bm q_T)   \tilde\rho_{Bc}(\bm k_T - \bm q_T)]}
{(q_T^2 + m^2)[(\bm k_T-\bm q_T)^2 + m^2]} f_{bca}
\right\},\label{eq:lalalalala_1}
\end{eqnarray}
and $a_{kai}^\dagger$ is the complex conjugate of~\eqref{eq:lalalalala_1}.

Finally, we now notice that we can further specify the 
choice of the gauge in which the glasma fields are computed. 
In fact, longitudinal invariance (or boost invariance in the case of 
a longitudinally expanding medium), 
as well as the condition $A_0=0$ (corresponding to
$A_\tau=0$, namely the Fock-Schwinger gauge, in the case of the
expanding system) imply that we can still perform gauge
transformations that involve the transverse plane
coordinates.
In particular, we can require that the fields satisfy the
Coulomb gauge condition, that is
\begin{equation}
q_i A^i_a(\bm q_T)=0.
\label{eq:coulomb_111_1005}
\end{equation}
We can achieve the condition~\eqref{eq:coulomb_111_1005} by applying
a transverse projector to the fields~\eqref{eq:lanostrasplendidachat}. When doing this, we notice that $A^i_a(\bm q_T) \propto q_i$,
implying that
they are purely longitudinal and the transverse projection gives zero.
Consequently, as long as we work in the Coulomb gauge~\eqref{eq:coulomb_111_1005},
we can neglect the fields~\eqref{eq:lanostrasplendidachat} in the computation
of the occupation numbers, that is in Eq.~\eqref{eq:lalalalala_1} we consider only the $z$ term.

\subsection{Calculation of $n_{kai}$\label{subsec:2}}

Next we turn to the longitudinal occupation numbers, obtained from Eq.~\eqref{eq:lalalalala_1} by putting $i=z$.
Taking into account that the fluctuations of the color charges on the two nuclei are uncorrelated, and using the Wick theorem to express the $4-$point correlator in terms of $2-$point correlators\footnote{The use of the Wick theorem in this context is justified by the fact that the fluctuations of the color charges are gaussian.}, we get
\begin{eqnarray}
\langle |a_{kaz}|^2 \rangle &=&
\frac{L_z^2}{2\omega_k}f_{abc}f_{ade} \int\frac{d^2q_T}{(2\pi)^2}
\int\frac{d^2\ell_T}{(2\pi)^2}
\frac{ \bm q_T\cdot (\bm k_T-\bm q_T)}
{(q_T^2 + m^2)((\bm k_T-\bm q_T)^2 + m^2)}
\frac{ \bm \ell_T\cdot (\bm k_T-\bm \ell_T) 
}
{(\ell_T^2 + m^2)((\bm k_T-\bm \ell_T)^2 + m^2)}\nonumber\\
&&\times
\langle\tilde\rho_{Ab}(\bm q_T)  \tilde\rho_{Ad}(-\bm \ell_T) \rangle  
\langle\tilde\rho_{Bc}(\bm k_T - \bm q_T)
 \tilde\rho_{Be}(\bm \ell_T - \bm k_T)\rangle.
\label{eq:pic_pot_1_1_94_50}
\end{eqnarray}
Notice that the correlators of the color charges in the right hand side of~\eqref{eq:pic_pot_1_1_94_50} entangle
the two momentum integrals.

The occupation numbers in
Eq.~\eqref{eq:pic_pot_1_1_94_50}
depend on the correlators of the color charges
that generate the glasma fields, see Eq.~\eqref{eq:rai2HD_00:26}.
In the transverse momentum space Eq.~\eqref{eq:rai2HD_00:26} reads
\begin{equation}
\langle \tilde \rho_a(\bm k_T)   \tilde \rho_b(\bm q_T)  \rangle =
\delta_{ab} f\left(\frac{\bm k_T - \bm q_T}{2}\right)
\int d^2v
\langle[g\mu(\bm v)]^2\rangle
e^{-i\bm v \cdot (\bm k_T + \bm q_T)},
\label{eq:rai2HD_00:34}
\end{equation}
where $f(\bm q)$ is the Fourier transform of $F$ in Eq.~\eqref{eq:rai2HD_00:26}.
In the spirit of the MV model, we assume that
color-charge fluctuations are uncorrelated in the transverse
plane, hence
$F(\bm u)$ in Eq.~\eqref{eq:rai2HD_00:26} is
\begin{equation}
F(\bm u) = \delta^2(\bm u).
\label{eq:F_u_def}
\end{equation}
Consequently $f(\bm u)=1$ and the momentum-space correlator
becomes
\begin{equation}
\langle \tilde \rho_a(\bm k_T)   \tilde \rho_b(\bm q_T)  \rangle =
\delta_{ab} 
\int d^2v
\langle[g\mu(\bm v)]^2\rangle
e^{-i\bm v \cdot (\bm k_T + \bm q_T)}.
\label{eq:rai2HD_00:34_2}
\end{equation}
Next we examine the correlator~\eqref{eq:rai2HD_00:34_2} in two
cases, namely for a coordinate-independent $\mu$ (which is a fair approximation to study the glasma fields produced in a small portion of a nucleus-nucleus collision), and for the coordinate-dependent $\mu$ used in hotspots models of the nucleons, that is closer to the actual implementations used for proton-nucleus collisions.
For the sake of simplicity, we dub the two cases as
AA- and pA-collisions respectively.

\end{widetext}

\subsubsection{MV model with coordinate-independent $\mu$: AA collisions}

\begin{figure}[t!]
\centering
\includegraphics[width=0.4\textwidth]{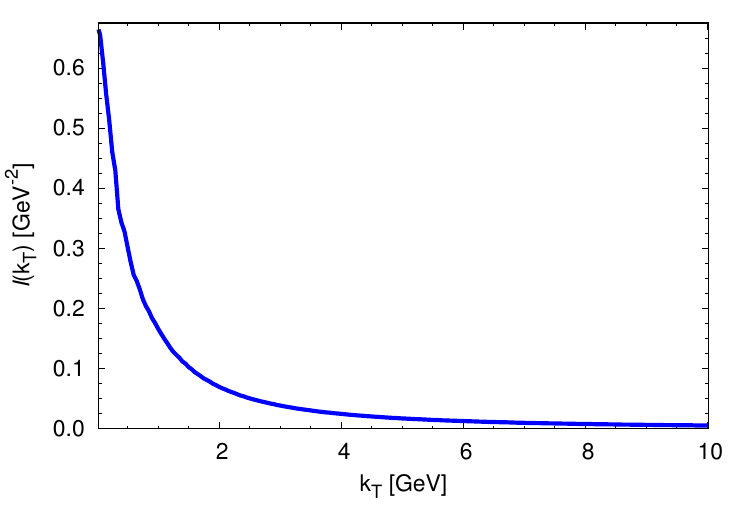}
\caption{The function $\mathcal{I}(k_T)$, Eq.~\eqref{eq:giallo_10_28}, versus $k_T$.}
\label{Fig:giallo_11_56}
\end{figure}

For a $\mu$ that does not depend on the
transverse plane coordinates, Eq.~\eqref{eq:rai2HD_00:34_2}
gives
\begin{equation}
\langle \tilde \rho_a(\bm k_T)   \tilde \rho_b(\bm q_T)  \rangle =
(2\pi)^2  (g\mu)^2 \delta_{ab} 
\delta^2 (\bm k_T + \bm q_T).
\label{eq:rai2HD_00:34_2_s}
\end{equation}
In this case, using $(2\pi)^2\delta^2(0)=A_T$, with $A_T=L_T^2$ 
corresponding to
the transverse area of the interaction region of two colliding nuclei, from \eqref{eq:pic_pot_1_1_94_50} we get
\begin{eqnarray}
\langle |a_{kaz}|^2 \rangle_{AA} &=&
\frac{3 L_z^2}{2\omega_k}
 (g\mu_A)^2(g\mu_B)^2A_T
\mathcal{I}(k_T),
\label{eq:giallo_10_14}
\end{eqnarray}
where we used
\begin{equation}
\sum_{b,c}|f_{abc}|^2 = \textcolor{black}{3}~~~~\mathrm{for~any~}a,
\label{eq:laresanoneunopzione}
\end{equation}
which can be easily verified. Moreover, we have defined
\begin{equation}
\mathcal{I}(k_T)=\int\frac{d^2\ell_T}{(2\pi)^2}
\frac{ [\bm \ell_T\cdot (\bm k_T-\bm \ell_T)]^2 
}
{(\ell_T^2 + m^2)^2((\bm k_T-\bm \ell_T)^2 + m^2)^2}.
\label{eq:giallo_10_28}
\end{equation}
From rotational invariance it follows that $\mathcal{I}$ depends on
the magnitude of $k_T$ only. 
For $\ell_T\gg k_T$ it is easily seen that the integrand 
in~\eqref{eq:laresanoneunopzione} behaves as
$\sim \ell_T^{-4}$, hence the integral is well defined in the UV.
Moreover, $m$ prevents any divergence in the
infrared domain.
In fact, for $k_T=0$ it is easy to check that
$\mathcal{I} = 1/12 \pi m^2$. The behavior of 
$\mathcal{I}(k_T)$ as a function of $k_T$ can be computed numerically and it is shown in Fig.~\ref{Fig:giallo_11_56} for $m=0.2$ GeV.

Taking into account Eq.~\eqref{eq:grandpa_shark}, 
as well as
\begin{equation}
d^3k=dk_x dk_y dk_z = \frac{(2\pi)^3}{L_z A_T},
\label{eq:rai_4_hd}
\end{equation}
we have
from Eq.~\eqref{eq:giallo_10_14}
\begin{equation}
n_{kaz}^{AA} = 
\frac{ 3(2\pi)^3}{\textcolor{black}{2}\omega_k}
 (g\mu_A)^2(g\mu_B)^2
\mathcal{I}(k_T)L_z.
\label{eq:giallo_12_09}
\end{equation}
It is useful to notice that the occupation numbers do not depend on $a$, in agreement with gauge invariance, hence, in the calculation of the entropy that we present in the next section,
the sum over $a$ simply gives an overall factor $N_c^2-1$.

\begin{figure}[t!]
\centering
\includegraphics[width=0.99\linewidth]{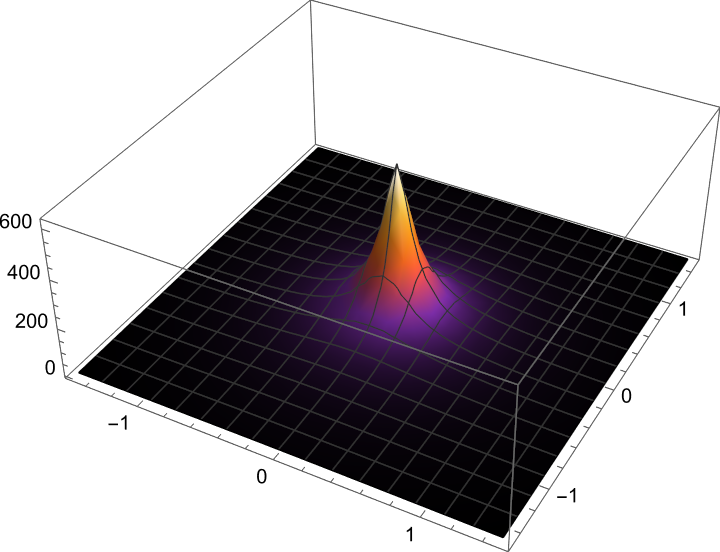}
\caption{
$n_{kaz}^{AA}$ defined in Eq.~\eqref{eq:giallo_12_09}, in the $(k_x,k_y)$ plane.
Calculations correspond to
$\mu_A = \mu_B = 0.5$ GeV
and $g=2$.
Units are GeV for $k_x$ and $k_y$.}
\label{Fig:consiglio_16_33}
\end{figure}

In Fig.~\ref{Fig:consiglio_16_33} we plot
$n_{kaz}^{AA}$ in the $(k_x,k_y)$ plane.
Calculations correspond to $\mu_A = \mu_B = 0.5$ GeV and $g=2$. We notice that the largest contribution to the occupation number comes from the modes with $|\bm k_T| < g\mu_{A/B}$.

\subsubsection{MV model with coordinate-dependent $\mu$: pA collisions}

If $\mu$ depends on the transverse plane coordinates, as it happens for the proton, we instead must consider
\begin{eqnarray}
&&\langle \rho_a(\bm x_T)   \rho_b(\bm y_T)  \rangle =
\langle\left[  g\mu \left(\bm v\right) \right]^2\rangle
 \delta_{ab} \delta^2(\bm u),
\label{eq:tv8HD_23:28}\\
&&\langle \tilde \rho_a(\bm k_T)   \tilde \rho_b(\bm q_T)  \rangle =
\delta_{ab}  \int d^2v
\langle[g\mu(\bm v)]^2\rangle
e^{-i\bm v \cdot (\bm k_T + \bm q_T)}.
\label{eq:tv8HD_23:28again}\nonumber\\
&&
\end{eqnarray}
The dependence of $\mu$ on coordinates, in the case of the proton, is given by~\cite{Schenke:2020mbo}
\begin{equation}
g\mu(\bm x_T) = \frac{c}{g}Q_s(\bm x_T),
\label{eq:tv8HD_23:59}
\end{equation}
where
\begin{equation}
Q_s^2(x,\bm x_T) = 
\frac{2\pi^2\alpha_s}{N_c}
xg(x,Q_0^2) T_p(\bm x_T).
\label{eq:loroeranopiuuniti}
\end{equation}
Here, $T_p(\bm x_T)$ denotes the thickness
function of the proton, defined as
\begin{equation}
T_p(\bm{x}_T) = 
\frac{1}{3} \sum_{i=1}^3 \frac{1}{2\pi B_q} \exp\left(-\frac{(\bm{x}_T-\bm{x}_i)^2}{2B_q}\right),
\label{eq:bd1}
\end{equation}
where $\bm x_i$ denote the positions of the constituent quarks, which are distributed
according to the gaussian distribution
\begin{equation}
T_{cq}(\bm{x}_i) = \frac{1}{2\pi B_{cq}}
\exp\left(-\frac{\bm{x}_i^2}{2B_{cq}}\right).
\label{eq:servono_rinforzi}
\end{equation}
Moreover, $xg(x,Q_0^2)$ in Eq.~\eqref{eq:loroeranopiuuniti} is the gluon distribution function at fixed $x$ and virtuality $Q_0^2$.
In this case, the ensemble average at a fixed value of $\bm x_T$ of the transverse
plane amounts to average $T_{p}(\bm x_T)$
over the locations of the constituent quarks~\cite{Parisi:2025slf}.  It is an easy exercise to show that
\begin{align}
\langle [Q_s (\bm v)]^2 \rangle &= \int d^2 x_i Q_s^2(\bm v) T_{cq}(\bm x_i)
\\
&=
\frac{2\pi\textcolor{black}{^2}\alpha_s}{N_c}(xg)\frac{e^{-v^2/2(B_q + B_{cq})}}{2\pi(B_q + B_{cq})},
\label{eq:rai2HD_00:49}
\end{align}
which gives
\begin{equation}
\langle [g\mu (\bm v)]^2 \rangle = 
\frac{c^2 \textcolor{black}{\pi}}{2N_c}(xg)\frac{e^{-v^2/2(B_q + B_{cq})}}{2\pi(B_q + B_{cq})}.
\label{eq:domanisera21:30}
\end{equation}
We also notice the useful relation
\begin{equation}
\int d^2v \langle [g\mu (\bm v)]^2 \rangle = 
\frac{c^2 \textcolor{black}{\pi}}{2N_c}(xg).
\label{eq:italia1_18:56}
\end{equation}
Using~\eqref{eq:domanisera21:30} in the correlators~\eqref{eq:tv8HD_23:28}
and~\eqref{eq:tv8HD_23:28again} we get
\begin{eqnarray}
&&\langle \rho_a(\bm x_T)   \rho_b(\bm y_T)  \rangle =
\frac{c^2 \textcolor{black}{\pi}}{2N_c}(xg)\frac{e^{-x_T^2/2(B_q + B_{cq})}}{2\pi(B_q + B_{cq})}
 \delta_{ab} \delta^2(\bm x_T - \bm y_T),
\nonumber\\
&&\label{eq:tv8HD_00:12}\\
&&\langle \tilde \rho_a(\bm k_T)   \tilde \rho_b(\bm q_T)  \rangle = 
\frac{c^2 \textcolor{black}{\pi}}{2N_c}(xg)
e^{-(\bm k_T + \bm q_T)^2(B_q + B_{cq})/2}\delta_{ab} .
\label{eq:tv8HD_00:12again}
\end{eqnarray}

By substituing in Eq.~\eqref{eq:pic_pot_1_1_94_50}, we easily get
\begin{equation} 
\langle |\alpha_{kaz}|^2 \rangle_{pA}  =
\frac{3L_z^2}{2\omega_k}
 (g\mu_A)^2  \frac{c^2\textcolor{black}{\pi}}{2N_c}(xg)
\mathcal{I}(k_T).
\label{eq:italia1_19_06}
\end{equation}
Incidentally, we notice that Eq.~\eqref{eq:italia1_19_06} can be
obtained from~\eqref{eq:giallo_10_14} by replacing
$(g\mu_B)^2 A_T$ with the ensemble-averaged value of the integral
of $(g\mu)^2$ on the transverse plane, see~\eqref{eq:italia1_18:56}.
Hence, the longitudinal occupation numbers are
\begin{equation}
n_{kaz}^{pA} = 
\frac{ 3(2\pi)^3}{\textcolor{black}{2}\omega_k}
 (g\mu_A)^2   \frac{c^2 \textcolor{black}{\pi}}{2N_c}(xg)   
\mathcal{I}(k_T)\frac{L_z}{A_T}.
\label{eq:italia1_19_01}
\end{equation}

\begin{figure}[t!]
\centering
\includegraphics[width=0.99\linewidth]{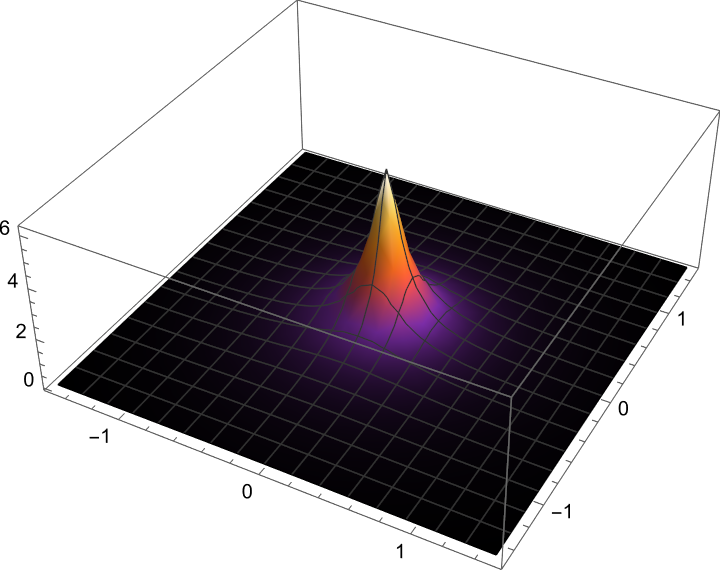}
\caption{
$n_{kaz}^{pA}$ defined in Eq.~\eqref{eq:italia1_19_01}, in the $(k_x,k_y)$ plane.
Units are GeV for $k_x$ and $k_y$.}
\label{Fig:italia1_20_46}
\end{figure}

In Fig.~\ref{Fig:italia1_20_46} we plot 
$n_{kaz}^{pA}$ 
in the $(k_x,k_y)$ plane. 
Results have been obtained with $c=1.25$ and $xg=3.94$, which are taken from~\cite{Schenke:2020mbo,Rezaeian:2012ji} and used also in~\cite{Parisi:2025slf}.
As expected, for the longitudinal numbers
we notice that the values for pA collisions are much smaller than the ones we find for AA collisions. This is due
to the fact that the contributions of the color charges of the nucleus and of the proton multiply each other to form the occupation numbers, hence the suppression of the charges of the proton causes also the suppression of the occupation number.

\section{Decoherence entropy and $S/N$\label{sec:entro_prod_a}}

In the previous section we have defined the occupation number of the coherent state of the Glasma within the MV model, for the
state $kai$, where $k$ labels momentum,
$a=1,\dots,N_c^2-1$ is the adjoint color index, and $i=x,y,z$. 

For the calculation of the decoherence entropy, it is useful to notice that for a density matrix given by the direct product of the density matrices for each quantum state $|kai\rangle$, that is
\begin{equation}
\rho = \prod_{kai}\rho_{kai},
\end{equation}
we can use
\begin{equation}
S = -\sum_{kai}\mathrm{Tr}(\rho_{kai}\log\rho_{kai})
=
-\sum_{kai}\sum_{\ell=1}^\infty \lambda_{kai,\ell}
\log \lambda_{kai,\ell},
\label{eq:unabusta}
\end{equation}
where $\lambda_{kai,\ell}$ is the $\ell-$th eigenvalue
of $\rho_{kai}$.

It is useful to summarize the results for the occupation numbers here, before we apply them to  the calculation of the decoherence entropy.
The occupation numbers of the coherent state explicitly depend on $k_T$ only, not on the color $a$. Moreover, the choice of the Coulomb gauge leaves only the contribution of the states with $i=z$.For this reason, we can simplify the notation and suppress the indices $a$ and $i$ from now on.

For AA collisions we have found
\begin{equation}
n_{k}^{AA} = 
\frac{ 3(2\pi)^3}{\textcolor{black}{2}\omega_k}
 (g\mu_A)^2(g\mu_B)^2
\mathcal{I}(k_T)L_z.
\label{eq:giallo_12_09_reprise}
\end{equation}
The integral $\mathcal{I}(k_T)$
is defined in Eq.~\eqref{eq:giallo_10_28}
and is finite both in the UV and in the IR.
Similarly, for the case of pA we have found
\begin{equation}
n_{k}^{pA} = 
\frac{ 3(2\pi)^3}{\textcolor{black}{2}\omega_k}
 (g\mu_A)^2   \frac{c^2 \textcolor{black}{\pi}}{2N_c}(xg)   
\mathcal{I}(k_T)\frac{L_z}{A_T}.
\label{eq:italia1_19_01_version2}
\end{equation}

The occupation numbers $n_{k}^{AA}$ and $n_{k}^{pA}$ are shown in Fig.~\ref{Fig:consiglio_16_33} and Fig.~\ref{Fig:italia1_20_46}, respectively.

Since we are interested in the decoherence entropy,
for the density operator we can use $\rho_\infty$
in Eq.~\eqref{eq:quantopesanoitrepuntididomanisera}.
In this case, for each coherent state
$kai$, the $\ell-th$ eigenvalue of the density operator is simply 
\begin{equation}
\lambda_{k,\ell} = e^{-n_{k}}\frac{n_{k}^\ell}
{\ell!},
%\label{eq:rigoresudumfriesNETTO}
\end{equation}
where we have considered that the occupation numbers are independent of $a$ and that only the states with $i=z$ contribute.
Consequently, the decoherence entropy can be written as
\begin{equation}
S_{\infty} = -
(N_c^2-1)\sum_{k_x,k_y}
\sum_\ell e^{-n_{k}}\frac{n_{k}^\ell}
{\ell!}\log \left(e^{-n_{k}}\frac{n_{k}^\ell}
{\ell!}\right).\label{eq:finale_1_1213}
\end{equation}
Here we have taken into account that the contribution is the same for $a=1,\dots,N_c^2-1$, hence the color degrees of freedom
appear as an overall degeneracy factor.
In~\eqref{eq:finale_1_1213} 
we will use Eqs.~\eqref{eq:giallo_12_09_reprise} 
and~\eqref{eq:italia1_19_01_version2} for AA and pA respectively, in place of $n_{k}$. Similarly,
the total particle number is
\begin{equation}
N = (N_c^2-1)\sum_{k_x,k_y} n_{k}.
\label{eq:total_number_1158}
\end{equation}

In the numerical calculations, we replace the summations over $(k_x,k_y)$
with an integration. By virtue of Eq.~\eqref{eq:rai_4_hd} we get 
\begin{equation}
dk_x dk_y  = \frac{(2\pi)^2}{A_T},
\label{eq:rai_4_hd?2D}
\end{equation}
hence we can write
\begin{equation}
N = (N_c^2-1)\frac{A_T}{(2\pi)^2}\int dk_x dk_y n_{k},
\label{eq:total_number_1158_conlim}
\end{equation}
and
\begin{align}
S_{\infty} &= -
(N_c^2-1)\frac{A_T}{(2\pi)^2} \nonumber \\
&\times\int dk_x dk_y
\sum_\ell e^{-n_{k}}\frac{n_{k}^\ell}
{\ell!}\log \left(e^{-n_{k}}\frac{n_{k}^\ell}
{\ell!}\right).\label{eq:finale_1_1213_contlim}
\end{align}

\begin{figure}[t!]
    \centering
    \includegraphics[width=0.4\textwidth]{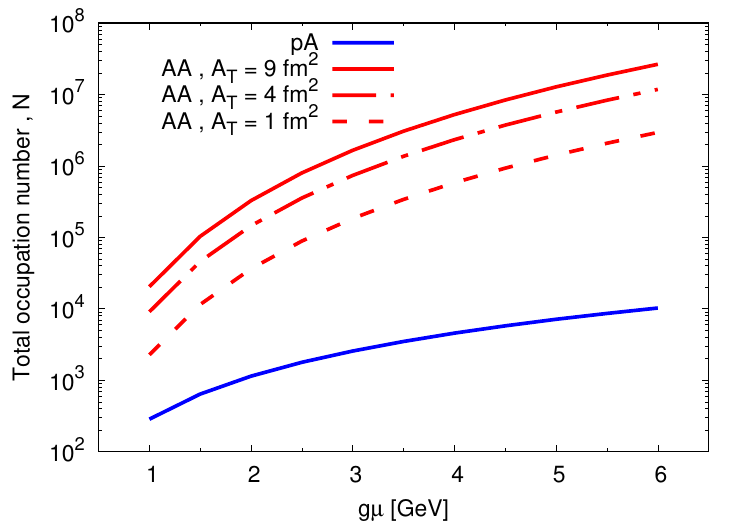}\\
        \includegraphics[width=0.4\textwidth]{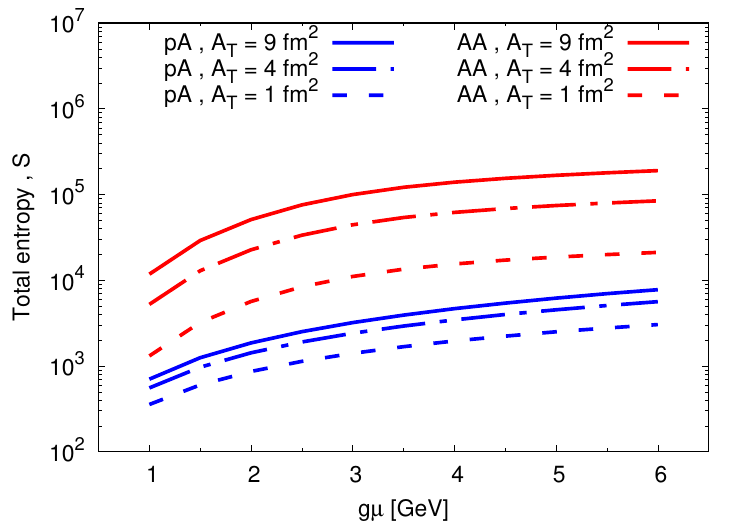}
    \caption{  Total occupation number $N$ (upper panel) and total decoherence entropy $S_{\infty}$ (lower panel) as function of energy parameter $g \mu$ for pA (blue) and AA (red) collisions. For $S_{\infty}$ the various lines show results for different values of transverse area $A_T$.}
    \label{fig:Smu1}
\end{figure}

In Fig.~\ref{fig:Smu1} we plot
$N$ (upper panel) and $S_{\infty}$ (lower panel) 
in pA and AA collisions according to Eqs.~\eqref{eq:total_number_1158_conlim} and \eqref{eq:finale_1_1213_contlim},  as function of $g \mu$. 
Values of $g\mu$ relevant for high-energy nuclear collisions are in the range $(1,3)$ GeV~\cite{Lappi:2007ku}, but for illustrative
purposes we plot the results up to $g\mu=6$ GeV.
Results are shown for three representative values of the transverse quantization area, $A_T$. 
The dependence of $N$ on $A_T$ for AA is
obviously related to the fact that in this case,
the system is homogeneous in the transverse plane; therefore, the occupation numbers scale linearly with $A_T$.
On the other hand, the occupation numbers in pA do not depend on $A_T$. This is also obvious, because in this case the interaction region is determined entirely by the distribution of the color charges in the proton.

Differently from what we have found for $N$,
$S_\infty$ has a dependence on $A_T$ both in the pA
and in the AA cases.
In particular, while in AA collisions the motivation is the same as for $N$, the dependence in the pA case is due to the fact that although $N$ does not depend on $A_T$, $n_{kaz}^{pA}$ does, see Eq.~\eqref{eq:italia1_19_01}, and the functional dependence of $S_\infty$ on $n_{kaz}^{pA}$ is non-trivial.
Therefore, some residual dependence on $A_T$ remains in $S_\infty$. Nevertheless, the $A_T$-dependence of $S_\infty$ in pA is milder than the one for AA.

\begin{figure}[t!]
    \centering
    \includegraphics[width=0.4\textwidth]{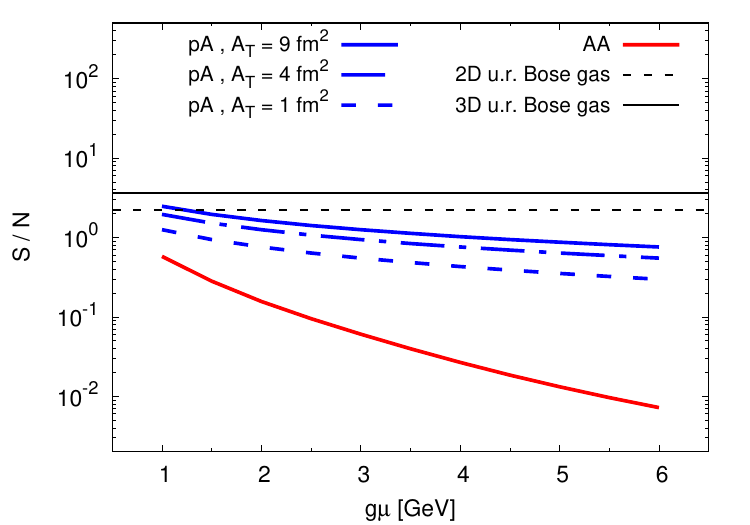} \\
    \includegraphics[width=0.4\textwidth]{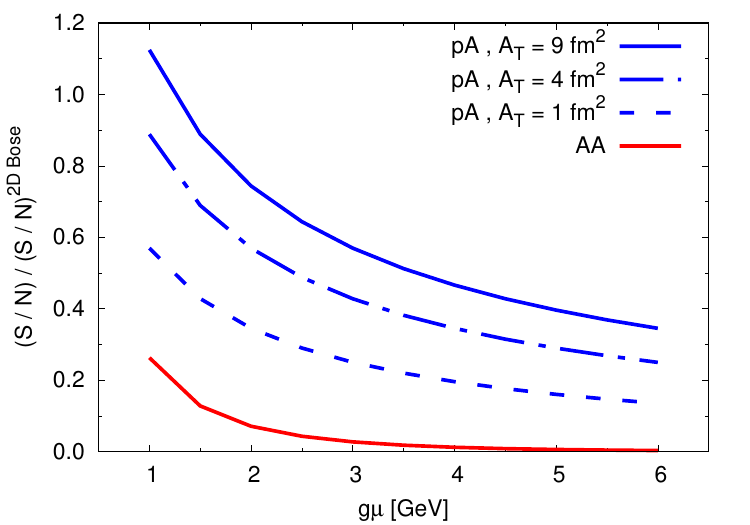}
    
    \caption{Upper panel: Entropy to number ratio $S/N$ as function of energy parameter $g \mu$ for pA (blue) and AA (red) collisions. For pA collisions various blue lines show results for different values of transverse area $A_T$. Since in AA collisions both $N$ and $S$ depend linearly on $A_T$, the ratio remains independent of it. The thin horizontal black lines represent the values of $S/N$ for an ultra-relativistic Bose gas in 2D (dashed) and 3D (solid). Lower panel: $S_{\infty}/N$ in pA (blue) and AA (red) are shown in linear scale and normalized 
    to the $S/N$ value for 2D the ultra-relativistic gluon gas.}
    \label{fig:ratiomu1}
\end{figure}

Figure~\ref{fig:ratiomu1} shows the main result of our work, namely the ratio $S_\infty/N$ 
for the coherent state that has undergone decoherence due to interactions with vacuum fluctuations.
The straight lines in  the upper panel of
Fig.~\ref{fig:ratiomu1}  represent the value of $S/N$
for a thermalized gas of gluons, that we denote
by $s_{SB}$. In particular,
the solid line corresponds to $s_{SB}$ for a three-dimensional gas, $s_{SB}=3.60$, while the dashed line stands for  a two-dimensional gas, $s_{SB}=2.19$. These results have been obtained using a standard
Bose-Einstein distribution for the gluons.
As we have assumed invariance along the longitudinal
direction, the problem at hand is effectively two-dimensional, therefore it meaningful to compare $S_{\infty}/N$
with the dashed straight line in the figure.  
This is done quantitatively in the lower panel of Fig.~\ref{fig:ratiomu1}, where the behavior of $S_{\infty}/N$
normalized to the value $s_{SB}=2.19$ from a two-dimensional 
ultra-relativistic gluon gas 
is shown as function of $g \mu$ in linear scale.

The results in Fig.~\ref{fig:ratiomu1} show interesting features.
Firstly, the results for AA are $A_T$-independent. This comes from the fact that both $N$ and $S_\infty$ scale linearly with
$A_T$ so the dependence drops when the ratio $S_\infty/N$
is considered. For this reason,
only one set of data is shown in the Figure for AA.
On the other hand, pA has a tiny dependence on $A_T$,
as a result of the behavior of $S_\infty$ discussed before.

We also notice that
$S_\infty/N$ for the decohered state is significantly 
below  $s_{SB}$, 
except for the pA case with the largest $A_T$ 
and the smallest $g\mu$. 
For example, for $g\mu = 2$ GeV, which represents a fair
value for a large nucleus in high-energy collisions at the LHC energy,
we find that $S_\infty/(N s_{SB})$
is $\sim 0.1$ for an AA collision, while it is
in the range $\sim(0.35,0.75)$ for a pA collision.
This indicates that in most cases
the decoherence alone drives the system relatively far from thermal equilibrium. This is in agreement with the picture drawn in~\cite{Iida:2014wea},
see in particular their Fig.~2, where it is shown that in an AA collision the vast majority of the entropy is produced at a later stage by instabilities.
It is however interesting that for pA collisions, 
the system after decoherence is closer to thermal equilibrium than the one produced in AA collisions.

Finally,
we notice that 
as $g\mu$ increases, the ratio $S/N$ 
moves increasingly away from the thermalized gas values.
Mathematically, this behavior corresponds to  
$N$ increasing faster than $S$ as $g\mu$ grows.
This trend can be qualitatively understood by noting that 
larger $g\mu$ implies a higher gluon density in the initial coherent state, which hence deviates further from a dilute gluon gas. 
Consequently, the decoherence process does not generate 
sufficient entropy to thermalize the system.

\section{Conclusions and outlook}
We have investigated the entropy production due to quantum decoherence of a coherent state modeled after the glasma fields generated in the early stages of high-energy nuclear collisions. 
By employing an analytical open quantum system framework within a  phase-damping model, we computed the asymptotic von Neumann entropy per particle, $S_\infty/N$, resulting from the interaction of the coherent state with vacuum fluctuations. 
The initial occupation numbers characterizing these coherent states were derived from a realistic description of proton-nucleus and nucleus-nucleus collisions, incorporating transverse spatial dependence via the McLerran-Venugopalan model.

Our results show that the entropy generated by decoherence alone is significantly below the values expected for a thermalized ultrarelativistic gluon gas, particularly in the nucleus-nucleus collision scenario. 
This indicates that quantum decoherence induced solely by vacuum fluctuations, as modeled here, 
is insufficient to fully thermalize the initial coherent state. 
In proton-nucleus collisions, the entropy per particle approaches the thermal limit only for small values of the glasma parameter $g\mu$ 
and for large transverse areas, but even in this regime decoherence alone cannot account for full thermalization.

Furthermore, we observe that as $g\mu$ increases, the ratio 
$S_\infty/N$ progressively diverges from the thermalized values. This trend can be qualitatively understood by the fact that higher $g\mu$ 
corresponds to an increased gluon density in the initial coherent state, which deviates further from a dilute gluon gas picture. Consequently, the occupation number $N$ 
grows faster than the entropy $S$, 
implying that decoherence-induced entropy production 
does not scale sufficiently to drive the system toward thermal equilibrium.

In the context of the pre-equilibrium
stage of high-energy nuclear collisions,
our study highlights the necessity to consider 
additional entropy-generating mechanisms beyond pure phase decoherence, 
such as non-abelian interactions with 
fluctuations that have to be added on top of the Glasma, which are expected to play a crucial role in the thermalization process. 
Incorporating these effects and extending the model to include amplitude-damping and non-Markovian dynamics represent important directions for future work.

\subsection*{Acknowledgments}
M. R. acknowledges Bruno Barbieri and John Petrucci for inspiration, and Giuseppe Chiatto for the numerous discussions on some of the topics presented in this article.
This work has been partly funded by the
European Union – Next Generation EU through the
research grant number P2022Z4P4B “SOPHYA - Sustainable Optimised PHYsics Algorithms: fundamental physics to build an advanced society” under the program PRIN 2022 PNRR of the Italian Ministero dell’Università
e Ricerca (MUR),
and by
PIACERI “Linea di intervento 1” (M@uRHIC) of the University of Catania. G.C. would like to thank Giuseppe Falci for fruitful discussions.
G. C. and S. P. acknowledge financial support from PNRR MUR Project PE0000023-NQSTI.

\subsection*{Data Availability Statement}
No datasets were generated or analyzed during the current study.

\bibliography{biblio-glasma-oqs.bib}

\begin{thebibliography}{58}%
\makeatletter
\providecommand \@ifxundefined [1]{%
 \@ifx{#1\undefined}
}%
\providecommand \@ifnum [1]{%
 \ifnum #1\expandafter \@firstoftwo
 \else \expandafter \@secondoftwo
 \fi
}%
\providecommand \@ifx [1]{%
 \ifx #1\expandafter \@firstoftwo
 \else \expandafter \@secondoftwo
 \fi
}%
\providecommand \natexlab [1]{#1}%
\providecommand \enquote  [1]{``#1''}%
\providecommand \bibnamefont  [1]{#1}%
\providecommand \bibfnamefont [1]{#1}%
\providecommand \citenamefont [1]{#1}%
\providecommand \href@noop [0]{\@secondoftwo}%
\providecommand \href [0]{\begingroup \@sanitize@url \@href}%
\providecommand \@href[1]{\@@startlink{#1}\@@href}%
\providecommand \@@href[1]{\endgroup#1\@@endlink}%
\providecommand \@sanitize@url [0]{\catcode `\\12\catcode `\$12\catcode `\&12\catcode `\#12\catcode `\^12\catcode `\_12\catcode `\%12\relax}%
\providecommand \@@startlink[1]{}%
\providecommand \@@endlink[0]{}%
\providecommand \url  [0]{\begingroup\@sanitize@url \@url }%
\providecommand \@url [1]{\endgroup\@href {#1}{\urlprefix }}%
\providecommand \urlprefix  [0]{URL }%
\providecommand \Eprint [0]{\href }%
\providecommand \doibase [0]{https://doi.org/}%
\providecommand \selectlanguage [0]{\@gobble}%
\providecommand \bibinfo  [0]{\@secondoftwo}%
\providecommand \bibfield  [0]{\@secondoftwo}%
\providecommand \translation [1]{[#1]}%
\providecommand \BibitemOpen [0]{}%
\providecommand \bibitemStop [0]{}%
\providecommand \bibitemNoStop [0]{.\EOS\space}%
\providecommand \EOS [0]{\spacefactor3000\relax}%
\providecommand \BibitemShut  [1]{\csname bibitem#1\endcsname}%
\let\auto@bib@innerbib\@empty
%</preamble>
\bibitem [{\citenamefont {Geiger}(1994)}]{Geiger:1994ip}%
  \BibitemOpen
  \bibfield  {author} {\bibinfo {author} {\bibfnamefont {K.}~\bibnamefont {Geiger}},\ }\bibfield  {title} {\bibinfo {title} {{High density QCD and entropy production at heavy ion colliders}},\ }in\ \href@noop {} {\emph {\bibinfo {booktitle} {{NATO Advanced Study Workshop on Hot Hadronic Matter: Theory and Experiment}}}}\ (\bibinfo {year} {1994})\ pp.\ \bibinfo {pages} {233--240},\ \Eprint {https://arxiv.org/abs/hep-ph/9409219} {arXiv:hep-ph/9409219} \BibitemShut {NoStop}%
\bibitem [{\citenamefont {Reiter}\ \emph {et~al.}(1998)\citenamefont {Reiter}, \citenamefont {Dumitru}, \citenamefont {Brachmann}, \citenamefont {Maruhn}, \citenamefont {Stoecker},\ and\ \citenamefont {Greiner}}]{Reiter:1998uq}%
  \BibitemOpen
  \bibfield  {author} {\bibinfo {author} {\bibfnamefont {M.}~\bibnamefont {Reiter}}, \bibinfo {author} {\bibfnamefont {A.}~\bibnamefont {Dumitru}}, \bibinfo {author} {\bibfnamefont {J.}~\bibnamefont {Brachmann}}, \bibinfo {author} {\bibfnamefont {J.~A.}\ \bibnamefont {Maruhn}}, \bibinfo {author} {\bibfnamefont {H.}~\bibnamefont {Stoecker}},\ and\ \bibinfo {author} {\bibfnamefont {W.}~\bibnamefont {Greiner}},\ }\bibfield  {title} {\bibinfo {title} {{Entropy production in collisions of relativistic heavy ions: A Signal for quark gluon plasma phase transition?}},\ }\href {https://doi.org/10.1016/S0375-9474(98)00556-9} {\bibfield  {journal} {\bibinfo  {journal} {Nucl. Phys. A}\ }\textbf {\bibinfo {volume} {643}},\ \bibinfo {pages} {99} (\bibinfo {year} {1998})},\ \Eprint {https://arxiv.org/abs/nucl-th/9806010} {arXiv:nucl-th/9806010} \BibitemShut {NoStop}%
\bibitem [{\citenamefont {Das}\ \emph {et~al.}(1986)\citenamefont {Das}, \citenamefont {Tripathi},\ and\ \citenamefont {Cugnon}}]{PhysRevLett.56.1663}%
  \BibitemOpen
  \bibfield  {author} {\bibinfo {author} {\bibfnamefont {C.}~\bibnamefont {Das}}, \bibinfo {author} {\bibfnamefont {R.~K.}\ \bibnamefont {Tripathi}},\ and\ \bibinfo {author} {\bibfnamefont {J.}~\bibnamefont {Cugnon}},\ }\bibfield  {title} {\bibinfo {title} {Entropy production in heavy-ion collisions},\ }\href {https://doi.org/10.1103/PhysRevLett.56.1663} {\bibfield  {journal} {\bibinfo  {journal} {Phys. Rev. Lett.}\ }\textbf {\bibinfo {volume} {56}},\ \bibinfo {pages} {1663} (\bibinfo {year} {1986})}\BibitemShut {NoStop}%
\bibitem [{\citenamefont {Kunihiro}\ \emph {et~al.}(2009)\citenamefont {Kunihiro}, \citenamefont {Muller}, \citenamefont {Ohnishi},\ and\ \citenamefont {Schafer}}]{Kunihiro:2008gv}%
  \BibitemOpen
  \bibfield  {author} {\bibinfo {author} {\bibfnamefont {T.}~\bibnamefont {Kunihiro}}, \bibinfo {author} {\bibfnamefont {B.}~\bibnamefont {Muller}}, \bibinfo {author} {\bibfnamefont {A.}~\bibnamefont {Ohnishi}},\ and\ \bibinfo {author} {\bibfnamefont {A.}~\bibnamefont {Schafer}},\ }\bibfield  {title} {\bibinfo {title} {{Towards a Theory of Entropy Production in the Little and Big Bang}},\ }\href {https://doi.org/10.1143/PTP.121.555} {\bibfield  {journal} {\bibinfo  {journal} {Prog. Theor. Phys.}\ }\textbf {\bibinfo {volume} {121}},\ \bibinfo {pages} {555} (\bibinfo {year} {2009})},\ \Eprint {https://arxiv.org/abs/0809.4831} {arXiv:0809.4831 [hep-ph]} \BibitemShut {NoStop}%
\bibitem [{\citenamefont {Fries}\ \emph {et~al.}(2009)\citenamefont {Fries}, \citenamefont {Kunihiro}, \citenamefont {Muller}, \citenamefont {Ohnishi},\ and\ \citenamefont {Schafer}}]{Fries:2009wh}%
  \BibitemOpen
  \bibfield  {author} {\bibinfo {author} {\bibfnamefont {R.~J.}\ \bibnamefont {Fries}}, \bibinfo {author} {\bibfnamefont {T.}~\bibnamefont {Kunihiro}}, \bibinfo {author} {\bibfnamefont {B.}~\bibnamefont {Muller}}, \bibinfo {author} {\bibfnamefont {A.}~\bibnamefont {Ohnishi}},\ and\ \bibinfo {author} {\bibfnamefont {A.}~\bibnamefont {Schafer}},\ }\bibfield  {title} {\bibinfo {title} {{From 0 to 5000 in 2 x 10**-24 seconds: Entropy production in relativistic heavy-ion collisions}},\ }\href {https://doi.org/10.1016/j.nuclphysa.2009.09.041} {\bibfield  {journal} {\bibinfo  {journal} {Nucl. Phys. A}\ }\textbf {\bibinfo {volume} {830}},\ \bibinfo {pages} {519C} (\bibinfo {year} {2009})},\ \Eprint {https://arxiv.org/abs/0906.5293} {arXiv:0906.5293 [nucl-th]} \BibitemShut {NoStop}%
\bibitem [{\citenamefont {Dumitru}\ \emph {et~al.}(2007)\citenamefont {Dumitru}, \citenamefont {Moln\'ar},\ and\ \citenamefont {Nara}}]{PhysRevC.76.024910}%
  \BibitemOpen
  \bibfield  {author} {\bibinfo {author} {\bibfnamefont {A.}~\bibnamefont {Dumitru}}, \bibinfo {author} {\bibfnamefont {E.}~\bibnamefont {Moln\'ar}},\ and\ \bibinfo {author} {\bibfnamefont {Y.}~\bibnamefont {Nara}},\ }\bibfield  {title} {\bibinfo {title} {Entropy production in high-energy heavy-ion collisions and the correlation of shear viscosity and thermalization time},\ }\href {https://doi.org/10.1103/PhysRevC.76.024910} {\bibfield  {journal} {\bibinfo  {journal} {Phys. Rev. C}\ }\textbf {\bibinfo {volume} {76}},\ \bibinfo {pages} {024910} (\bibinfo {year} {2007})}\BibitemShut {NoStop}%
\bibitem [{\citenamefont {Ivanov}\ and\ \citenamefont {Soldatov}(2016)}]{Ivanov:2016hes}%
  \BibitemOpen
  \bibfield  {author} {\bibinfo {author} {\bibfnamefont {Y.~B.}\ \bibnamefont {Ivanov}}\ and\ \bibinfo {author} {\bibfnamefont {A.~A.}\ \bibnamefont {Soldatov}},\ }\bibfield  {title} {\bibinfo {title} {{Entropy Production and Effective Viscosity in Heavy-Ion Collisions}},\ }\href {https://doi.org/10.1140/epja/i2016-16367-7} {\bibfield  {journal} {\bibinfo  {journal} {Eur. Phys. J. A}\ }\textbf {\bibinfo {volume} {52}},\ \bibinfo {pages} {367} (\bibinfo {year} {2016})},\ \Eprint {https://arxiv.org/abs/1605.02476} {arXiv:1605.02476 [nucl-th]} \BibitemShut {NoStop}%
\bibitem [{\citenamefont {Muller}\ and\ \citenamefont {Schafer}(2011)}]{Muller:2011ra}%
  \BibitemOpen
  \bibfield  {author} {\bibinfo {author} {\bibfnamefont {B.}~\bibnamefont {Muller}}\ and\ \bibinfo {author} {\bibfnamefont {A.}~\bibnamefont {Schafer}},\ }\bibfield  {title} {\bibinfo {title} {{Entropy Creation in Relativistic Heavy Ion Collisions}},\ }\href {https://doi.org/10.1142/S0218301311020459} {\bibfield  {journal} {\bibinfo  {journal} {Int. J. Mod. Phys. E}\ }\textbf {\bibinfo {volume} {20}},\ \bibinfo {pages} {2235} (\bibinfo {year} {2011})},\ \Eprint {https://arxiv.org/abs/1110.2378} {arXiv:1110.2378 [hep-ph]} \BibitemShut {NoStop}%
\bibitem [{\citenamefont {Zurek}\ and\ \citenamefont {Paz}(1994)}]{Zurek:1994wd}%
  \BibitemOpen
  \bibfield  {author} {\bibinfo {author} {\bibfnamefont {W.~H.}\ \bibnamefont {Zurek}}\ and\ \bibinfo {author} {\bibfnamefont {J.~P.}\ \bibnamefont {Paz}},\ }\bibfield  {title} {\bibinfo {title} {{Decoherence, chaos, and the second law}},\ }\href {https://doi.org/10.1103/PhysRevLett.72.2508} {\bibfield  {journal} {\bibinfo  {journal} {Phys. Rev. Lett.}\ }\textbf {\bibinfo {volume} {72}},\ \bibinfo {pages} {2508} (\bibinfo {year} {1994})},\ \Eprint {https://arxiv.org/abs/gr-qc/9402006} {arXiv:gr-qc/9402006} \BibitemShut {NoStop}%
\bibitem [{\citenamefont {Elze}(1995)}]{Elze:1994qa}%
  \BibitemOpen
  \bibfield  {author} {\bibinfo {author} {\bibfnamefont {H.-T.}\ \bibnamefont {Elze}},\ }\bibfield  {title} {\bibinfo {title} {{Quantum decoherence, entropy and thermalization in strong interactions at high-energy. 1: Noisy and dissipative vacuum effects in toy models}},\ }\href {https://doi.org/10.1016/0550-3213(94)00523-H} {\bibfield  {journal} {\bibinfo  {journal} {Nucl. Phys. B}\ }\textbf {\bibinfo {volume} {436}},\ \bibinfo {pages} {213} (\bibinfo {year} {1995})},\ \Eprint {https://arxiv.org/abs/hep-ph/9404215} {arXiv:hep-ph/9404215} \BibitemShut {NoStop}%
\bibitem [{\citenamefont {Rais}\ \emph {et~al.}(2025)\citenamefont {Rais}, \citenamefont {van Hees},\ and\ \citenamefont {Greiner}}]{Rais:2025fps}%
  \BibitemOpen
  \bibfield  {author} {\bibinfo {author} {\bibfnamefont {J.}~\bibnamefont {Rais}}, \bibinfo {author} {\bibfnamefont {H.}~\bibnamefont {van Hees}},\ and\ \bibinfo {author} {\bibfnamefont {C.}~\bibnamefont {Greiner}},\ }\bibfield  {title} {\bibinfo {title} {{Bound-state formation and thermalization within the Lindblad approach}},\ }\href {https://doi.org/10.1103/PhysRevC.111.054918} {\bibfield  {journal} {\bibinfo  {journal} {Phys. Rev. C}\ }\textbf {\bibinfo {volume} {111}},\ \bibinfo {pages} {054918} (\bibinfo {year} {2025})}\BibitemShut {NoStop}%
\bibitem [{\citenamefont {Delorme}\ \emph {et~al.}(2024)\citenamefont {Delorme}, \citenamefont {Katz}, \citenamefont {Gousset}, \citenamefont {Gossiaux},\ and\ \citenamefont {Blaizot}}]{Delorme:2024rdo}%
  \BibitemOpen
  \bibfield  {author} {\bibinfo {author} {\bibfnamefont {S.}~\bibnamefont {Delorme}}, \bibinfo {author} {\bibfnamefont {R.}~\bibnamefont {Katz}}, \bibinfo {author} {\bibfnamefont {T.}~\bibnamefont {Gousset}}, \bibinfo {author} {\bibfnamefont {P.~B.}\ \bibnamefont {Gossiaux}},\ and\ \bibinfo {author} {\bibfnamefont {J.-P.}\ \bibnamefont {Blaizot}},\ }\bibfield  {title} {\bibinfo {title} {{Quarkonium dynamics in the quantum Brownian regime with non-abelian quantum master equations}},\ }\href {https://doi.org/10.1007/JHEP06(2024)060} {\bibfield  {journal} {\bibinfo  {journal} {JHEP}\ }\textbf {\bibinfo {volume} {06}},\ \bibinfo {pages} {060}},\ \Eprint {https://arxiv.org/abs/2402.04488} {arXiv:2402.04488 [hep-ph]} \BibitemShut {NoStop}%
\bibitem [{\citenamefont {Neidig}\ \emph {et~al.}(2024)\citenamefont {Neidig}, \citenamefont {Rais}, \citenamefont {Bleicher}, \citenamefont {van Hees},\ and\ \citenamefont {Greiner}}]{Neidig:2023kid}%
  \BibitemOpen
  \bibfield  {author} {\bibinfo {author} {\bibfnamefont {T.}~\bibnamefont {Neidig}}, \bibinfo {author} {\bibfnamefont {J.}~\bibnamefont {Rais}}, \bibinfo {author} {\bibfnamefont {M.}~\bibnamefont {Bleicher}}, \bibinfo {author} {\bibfnamefont {H.}~\bibnamefont {van Hees}},\ and\ \bibinfo {author} {\bibfnamefont {C.}~\bibnamefont {Greiner}},\ }\bibfield  {title} {\bibinfo {title} {{Open quantum systems with Kadanoff-Baym equations}},\ }\href {https://doi.org/10.1016/j.physletb.2024.138589} {\bibfield  {journal} {\bibinfo  {journal} {Phys. Lett. B}\ }\textbf {\bibinfo {volume} {851}},\ \bibinfo {pages} {138589} (\bibinfo {year} {2024})},\ \Eprint {https://arxiv.org/abs/2308.07659} {arXiv:2308.07659 [nucl-th]} \BibitemShut {NoStop}%
\bibitem [{\citenamefont {Brambilla}\ \emph {et~al.}(2021)\citenamefont {Brambilla}, \citenamefont {Escobedo}, \citenamefont {Strickland}, \citenamefont {Vairo}, \citenamefont {Vander~Griend},\ and\ \citenamefont {Weber}}]{Brambilla:2020qwo}%
  \BibitemOpen
  \bibfield  {author} {\bibinfo {author} {\bibfnamefont {N.}~\bibnamefont {Brambilla}}, \bibinfo {author} {\bibfnamefont {M.~A.}\ \bibnamefont {Escobedo}}, \bibinfo {author} {\bibfnamefont {M.}~\bibnamefont {Strickland}}, \bibinfo {author} {\bibfnamefont {A.}~\bibnamefont {Vairo}}, \bibinfo {author} {\bibfnamefont {P.}~\bibnamefont {Vander~Griend}},\ and\ \bibinfo {author} {\bibfnamefont {J.~H.}\ \bibnamefont {Weber}},\ }\bibfield  {title} {\bibinfo {title} {{Bottomonium suppression in an open quantum system using the quantum trajectories method}},\ }\href {https://doi.org/10.1007/JHEP05(2021)136} {\bibfield  {journal} {\bibinfo  {journal} {JHEP}\ }\textbf {\bibinfo {volume} {05}},\ \bibinfo {pages} {136}},\ \Eprint {https://arxiv.org/abs/2012.01240} {arXiv:2012.01240 [hep-ph]} \BibitemShut {NoStop}%
\bibitem [{\citenamefont {Kajimoto}\ \emph {et~al.}(2018)\citenamefont {Kajimoto}, \citenamefont {Akamatsu}, \citenamefont {Asakawa},\ and\ \citenamefont {Rothkopf}}]{Kajimoto:2017rel}%
  \BibitemOpen
  \bibfield  {author} {\bibinfo {author} {\bibfnamefont {S.}~\bibnamefont {Kajimoto}}, \bibinfo {author} {\bibfnamefont {Y.}~\bibnamefont {Akamatsu}}, \bibinfo {author} {\bibfnamefont {M.}~\bibnamefont {Asakawa}},\ and\ \bibinfo {author} {\bibfnamefont {A.}~\bibnamefont {Rothkopf}},\ }\bibfield  {title} {\bibinfo {title} {{Dynamical dissociation of quarkonia by wave function decoherence}},\ }\href {https://doi.org/10.1103/PhysRevD.97.014003} {\bibfield  {journal} {\bibinfo  {journal} {Phys. Rev. D}\ }\textbf {\bibinfo {volume} {97}},\ \bibinfo {pages} {014003} (\bibinfo {year} {2018})},\ \Eprint {https://arxiv.org/abs/1705.03365} {arXiv:1705.03365 [nucl-th]} \BibitemShut {NoStop}%
\bibitem [{\citenamefont {Brambilla}\ \emph {et~al.}(2017)\citenamefont {Brambilla}, \citenamefont {Escobedo}, \citenamefont {Soto},\ and\ \citenamefont {Vairo}}]{Brambilla:2016wgg}%
  \BibitemOpen
  \bibfield  {author} {\bibinfo {author} {\bibfnamefont {N.}~\bibnamefont {Brambilla}}, \bibinfo {author} {\bibfnamefont {M.~A.}\ \bibnamefont {Escobedo}}, \bibinfo {author} {\bibfnamefont {J.}~\bibnamefont {Soto}},\ and\ \bibinfo {author} {\bibfnamefont {A.}~\bibnamefont {Vairo}},\ }\bibfield  {title} {\bibinfo {title} {{Quarkonium suppression in heavy-ion collisions: an open quantum system approach}},\ }\href {https://doi.org/10.1103/PhysRevD.96.034021} {\bibfield  {journal} {\bibinfo  {journal} {Phys. Rev. D}\ }\textbf {\bibinfo {volume} {96}},\ \bibinfo {pages} {034021} (\bibinfo {year} {2017})},\ \Eprint {https://arxiv.org/abs/1612.07248} {arXiv:1612.07248 [hep-ph]} \BibitemShut {NoStop}%
\bibitem [{\citenamefont {Akamatsu}\ and\ \citenamefont {Rothkopf}(2012)}]{Akamatsu:2011se}%
  \BibitemOpen
  \bibfield  {author} {\bibinfo {author} {\bibfnamefont {Y.}~\bibnamefont {Akamatsu}}\ and\ \bibinfo {author} {\bibfnamefont {A.}~\bibnamefont {Rothkopf}},\ }\bibfield  {title} {\bibinfo {title} {{Stochastic potential and quantum decoherence of heavy quarkonium in the quark-gluon plasma}},\ }\href {https://doi.org/10.1103/PhysRevD.85.105011} {\bibfield  {journal} {\bibinfo  {journal} {Phys. Rev. D}\ }\textbf {\bibinfo {volume} {85}},\ \bibinfo {pages} {105011} (\bibinfo {year} {2012})},\ \Eprint {https://arxiv.org/abs/1110.1203} {arXiv:1110.1203 [hep-ph]} \BibitemShut {NoStop}%
\bibitem [{\citenamefont {Blaizot}\ and\ \citenamefont {Escobedo}(2018)}]{Blaizot:2017ypk}%
  \BibitemOpen
  \bibfield  {author} {\bibinfo {author} {\bibfnamefont {J.-P.}\ \bibnamefont {Blaizot}}\ and\ \bibinfo {author} {\bibfnamefont {M.~A.}\ \bibnamefont {Escobedo}},\ }\bibfield  {title} {\bibinfo {title} {{Quantum and classical dynamics of heavy quarks in a quark-gluon plasma}},\ }\href {https://doi.org/10.1007/JHEP06(2018)034} {\bibfield  {journal} {\bibinfo  {journal} {JHEP}\ }\textbf {\bibinfo {volume} {06}},\ \bibinfo {pages} {034}},\ \Eprint {https://arxiv.org/abs/1711.10812} {arXiv:1711.10812 [hep-ph]} \BibitemShut {NoStop}%
\bibitem [{\citenamefont {Katz}\ and\ \citenamefont {Gossiaux}(2016)}]{Katz:2015qja}%
  \BibitemOpen
  \bibfield  {author} {\bibinfo {author} {\bibfnamefont {R.}~\bibnamefont {Katz}}\ and\ \bibinfo {author} {\bibfnamefont {P.~B.}\ \bibnamefont {Gossiaux}},\ }\bibfield  {title} {\bibinfo {title} {{The Schr\"odinger\textendash{}Langevin equation with and without thermal fluctuations}},\ }\href {https://doi.org/10.1016/j.aop.2016.02.005} {\bibfield  {journal} {\bibinfo  {journal} {Annals Phys.}\ }\textbf {\bibinfo {volume} {368}},\ \bibinfo {pages} {267} (\bibinfo {year} {2016})},\ \Eprint {https://arxiv.org/abs/1504.08087} {arXiv:1504.08087 [quant-ph]} \BibitemShut {NoStop}%
\bibitem [{\citenamefont {De~Jong}\ \emph {et~al.}(2021)\citenamefont {De~Jong}, \citenamefont {Metcalf}, \citenamefont {Mulligan}, \citenamefont {P{\l}osko{\'n}}, \citenamefont {Ringer},\ and\ \citenamefont {Yao}}]{DeJong:2020riy}%
  \BibitemOpen
  \bibfield  {author} {\bibinfo {author} {\bibfnamefont {W.~A.}\ \bibnamefont {De~Jong}}, \bibinfo {author} {\bibfnamefont {M.}~\bibnamefont {Metcalf}}, \bibinfo {author} {\bibfnamefont {J.}~\bibnamefont {Mulligan}}, \bibinfo {author} {\bibfnamefont {M.}~\bibnamefont {P{\l}osko{\'n}}}, \bibinfo {author} {\bibfnamefont {F.}~\bibnamefont {Ringer}},\ and\ \bibinfo {author} {\bibfnamefont {X.}~\bibnamefont {Yao}},\ }\bibfield  {title} {\bibinfo {title} {{Quantum simulation of open quantum systems in heavy-ion collisions}},\ }\href {https://doi.org/10.1103/PhysRevD.104.L051501} {\bibfield  {journal} {\bibinfo  {journal} {Phys. Rev. D}\ }\textbf {\bibinfo {volume} {104}},\ \bibinfo {pages} {051501} (\bibinfo {year} {2021})},\ \Eprint {https://arxiv.org/abs/2010.03571} {arXiv:2010.03571 [hep-ph]} \BibitemShut {NoStop}%
\bibitem [{\citenamefont {Lindblad}(1976)}]{Lindblad:1975ef}%
  \BibitemOpen
  \bibfield  {author} {\bibinfo {author} {\bibfnamefont {G.}~\bibnamefont {Lindblad}},\ }\bibfield  {title} {\bibinfo {title} {{On the Generators of Quantum Dynamical Semigroups}},\ }\href {https://doi.org/10.1007/BF01608499} {\bibfield  {journal} {\bibinfo  {journal} {Commun. Math. Phys.}\ }\textbf {\bibinfo {volume} {48}},\ \bibinfo {pages} {119} (\bibinfo {year} {1976})}\BibitemShut {NoStop}%
\bibitem [{\citenamefont {Gorini}\ \emph {et~al.}(1976)\citenamefont {Gorini}, \citenamefont {Kossakowski},\ and\ \citenamefont {Sudarshan}}]{Gorini:1975nb}%
  \BibitemOpen
  \bibfield  {author} {\bibinfo {author} {\bibfnamefont {V.}~\bibnamefont {Gorini}}, \bibinfo {author} {\bibfnamefont {A.}~\bibnamefont {Kossakowski}},\ and\ \bibinfo {author} {\bibfnamefont {E.~C.~G.}\ \bibnamefont {Sudarshan}},\ }\bibfield  {title} {\bibinfo {title} {{Completely Positive Dynamical Semigroups of N Level Systems}},\ }\href {https://doi.org/10.1063/1.522979} {\bibfield  {journal} {\bibinfo  {journal} {J. Math. Phys.}\ }\textbf {\bibinfo {volume} {17}},\ \bibinfo {pages} {821} (\bibinfo {year} {1976})}\BibitemShut {NoStop}%
\bibitem [{\citenamefont {Carmichael}(1999)}]{Carmichael::385006}%
  \BibitemOpen
  \bibfield  {author} {\bibinfo {author} {\bibfnamefont {H.~J.}\ \bibnamefont {Carmichael}},\ }\href {https://bib-pubdb1.desy.de/record/385006} {\emph {\bibinfo {title} {{S}tatistical methods in quantum optics: {V}ol. 1: {M}aster equations and {F}okker-{P}lanck equations; 1st ed., 2nd corr. print}}},\ Texts and monographs in physics\ (\bibinfo  {publisher} {Springer},\ \bibinfo {address} {Berlin},\ \bibinfo {year} {1999})\ p.\ \bibinfo {pages} {365 p}\BibitemShut {NoStop}%
\bibitem [{\citenamefont {Breuer}\ and\ \citenamefont {Petruccione}(2007)}]{Breuer:2007juk}%
  \BibitemOpen
  \bibfield  {author} {\bibinfo {author} {\bibfnamefont {H.-P.}\ \bibnamefont {Breuer}}\ and\ \bibinfo {author} {\bibfnamefont {F.}~\bibnamefont {Petruccione}},\ }\href {https://doi.org/10.1093/acprof:oso/9780199213900.001.0001} {\emph {\bibinfo {title} {{The Theory of Open Quantum Systems}}}}\ (\bibinfo  {publisher} {Oxford University Press},\ \bibinfo {year} {2007})\BibitemShut {NoStop}%
\bibitem [{\citenamefont {Vidiella-Barranco}(2016)}]{Vidiella-Barranco:2016hnh}%
  \BibitemOpen
  \bibfield  {author} {\bibinfo {author} {\bibfnamefont {A.}~\bibnamefont {Vidiella-Barranco}},\ }\bibfield  {title} {\bibinfo {title} {{Evolution of a quantum harmonic oscillator coupled to a minimal thermal environment}},\ }\href {https://doi.org/10.1016/j.physa.2016.04.033} {\bibfield  {journal} {\bibinfo  {journal} {Physica A}\ }\textbf {\bibinfo {volume} {459}},\ \bibinfo {pages} {78} (\bibinfo {year} {2016})},\ \Eprint {https://arxiv.org/abs/1605.01050} {arXiv:1605.01050 [quant-ph]} \BibitemShut {NoStop}%
\bibitem [{\citenamefont {Estes}\ \emph {et~al.}(1968)\citenamefont {Estes}, \citenamefont {Keil},\ and\ \citenamefont {Narducci}}]{Estes:1968quantum}%
  \BibitemOpen
  \bibfield  {author} {\bibinfo {author} {\bibfnamefont {L.~E.}\ \bibnamefont {Estes}}, \bibinfo {author} {\bibfnamefont {T.~H.}\ \bibnamefont {Keil}},\ and\ \bibinfo {author} {\bibfnamefont {L.~M.}\ \bibnamefont {Narducci}},\ }\bibfield  {title} {\bibinfo {title} {Quantum-mechanical description of two coupled harmonic oscillators},\ }\href@noop {} {\bibfield  {journal} {\bibinfo  {journal} {Physical Review}\ }\textbf {\bibinfo {volume} {175}},\ \bibinfo {pages} {286} (\bibinfo {year} {1968})}\BibitemShut {NoStop}%
\bibitem [{\citenamefont {Davies}(1976)}]{Davies:1976quantum}%
  \BibitemOpen
  \bibfield  {author} {\bibinfo {author} {\bibfnamefont {E.~B.}\ \bibnamefont {Davies}},\ }\href@noop {} {\emph {\bibinfo {title} {Quantum Theory of Open Systems}}}\ (\bibinfo  {publisher} {Academic Press},\ \bibinfo {year} {1976})\BibitemShut {NoStop}%
\bibitem [{\citenamefont {Mandel}\ and\ \citenamefont {Wolf}(1995)}]{Mandel:1995optical}%
  \BibitemOpen
  \bibfield  {author} {\bibinfo {author} {\bibfnamefont {L.}~\bibnamefont {Mandel}}\ and\ \bibinfo {author} {\bibfnamefont {E.}~\bibnamefont {Wolf}},\ }\href@noop {} {\emph {\bibinfo {title} {Optical Coherence and Quantum Optics}}}\ (\bibinfo  {publisher} {Cambridge University Press},\ \bibinfo {year} {1995})\BibitemShut {NoStop}%
\bibitem [{\citenamefont {McLerran}\ and\ \citenamefont {Venugopalan}(1994{\natexlab{a}})}]{McLerran:1993ni}%
  \BibitemOpen
  \bibfield  {author} {\bibinfo {author} {\bibfnamefont {L.~D.}\ \bibnamefont {McLerran}}\ and\ \bibinfo {author} {\bibfnamefont {R.}~\bibnamefont {Venugopalan}},\ }\bibfield  {title} {\bibinfo {title} {{Computing quark and gluon distribution functions for very large nuclei}},\ }\href {https://doi.org/10.1103/PhysRevD.49.2233} {\bibfield  {journal} {\bibinfo  {journal} {Phys. Rev. D}\ }\textbf {\bibinfo {volume} {49}},\ \bibinfo {pages} {2233} (\bibinfo {year} {1994}{\natexlab{a}})},\ \Eprint {https://arxiv.org/abs/hep-ph/9309289} {arXiv:hep-ph/9309289} \BibitemShut {NoStop}%
\bibitem [{\citenamefont {McLerran}\ and\ \citenamefont {Venugopalan}(1994{\natexlab{b}})}]{McLerran:1993ka}%
  \BibitemOpen
  \bibfield  {author} {\bibinfo {author} {\bibfnamefont {L.~D.}\ \bibnamefont {McLerran}}\ and\ \bibinfo {author} {\bibfnamefont {R.}~\bibnamefont {Venugopalan}},\ }\bibfield  {title} {\bibinfo {title} {{Gluon distribution functions for very large nuclei at small transverse momentum}},\ }\href {https://doi.org/10.1103/PhysRevD.49.3352} {\bibfield  {journal} {\bibinfo  {journal} {Phys. Rev. D}\ }\textbf {\bibinfo {volume} {49}},\ \bibinfo {pages} {3352} (\bibinfo {year} {1994}{\natexlab{b}})},\ \Eprint {https://arxiv.org/abs/hep-ph/9311205} {arXiv:hep-ph/9311205} \BibitemShut {NoStop}%
\bibitem [{\citenamefont {McLerran}\ and\ \citenamefont {Venugopalan}(1994{\natexlab{c}})}]{McLerran:1994vd}%
  \BibitemOpen
  \bibfield  {author} {\bibinfo {author} {\bibfnamefont {L.~D.}\ \bibnamefont {McLerran}}\ and\ \bibinfo {author} {\bibfnamefont {R.}~\bibnamefont {Venugopalan}},\ }\bibfield  {title} {\bibinfo {title} {{Green's functions in the color field of a large nucleus}},\ }\href {https://doi.org/10.1103/PhysRevD.50.2225} {\bibfield  {journal} {\bibinfo  {journal} {Phys. Rev. D}\ }\textbf {\bibinfo {volume} {50}},\ \bibinfo {pages} {2225} (\bibinfo {year} {1994}{\natexlab{c}})},\ \Eprint {https://arxiv.org/abs/hep-ph/9402335} {arXiv:hep-ph/9402335} \BibitemShut {NoStop}%
\bibitem [{\citenamefont {Iancu}\ \emph {et~al.}(2001)\citenamefont {Iancu}, \citenamefont {Leonidov},\ and\ \citenamefont {McLerran}}]{Iancu:2000hn}%
  \BibitemOpen
  \bibfield  {author} {\bibinfo {author} {\bibfnamefont {E.}~\bibnamefont {Iancu}}, \bibinfo {author} {\bibfnamefont {A.}~\bibnamefont {Leonidov}},\ and\ \bibinfo {author} {\bibfnamefont {L.~D.}\ \bibnamefont {McLerran}},\ }\bibfield  {title} {\bibinfo {title} {{Nonlinear gluon evolution in the color glass condensate. 1.}},\ }\href {https://doi.org/10.1016/S0375-9474(01)00642-X} {\bibfield  {journal} {\bibinfo  {journal} {Nucl. Phys. A}\ }\textbf {\bibinfo {volume} {692}},\ \bibinfo {pages} {583} (\bibinfo {year} {2001})},\ \Eprint {https://arxiv.org/abs/hep-ph/0011241} {arXiv:hep-ph/0011241} \BibitemShut {NoStop}%
\bibitem [{\citenamefont {Fukushima}\ and\ \citenamefont {Gelis}(2012)}]{Fukushima:2011nq}%
  \BibitemOpen
  \bibfield  {author} {\bibinfo {author} {\bibfnamefont {K.}~\bibnamefont {Fukushima}}\ and\ \bibinfo {author} {\bibfnamefont {F.}~\bibnamefont {Gelis}},\ }\bibfield  {title} {\bibinfo {title} {{The evolving Glasma}},\ }\href {https://doi.org/10.1016/j.nuclphysa.2011.11.003} {\bibfield  {journal} {\bibinfo  {journal} {Nucl. Phys. A}\ }\textbf {\bibinfo {volume} {874}},\ \bibinfo {pages} {108} (\bibinfo {year} {2012})},\ \Eprint {https://arxiv.org/abs/1106.1396} {arXiv:1106.1396 [hep-ph]} \BibitemShut {NoStop}%
\bibitem [{\citenamefont {Gelis}\ \emph {et~al.}(2010)\citenamefont {Gelis}, \citenamefont {Iancu}, \citenamefont {Jalilian-Marian},\ and\ \citenamefont {Venugopalan}}]{Gelis:2010nm}%
  \BibitemOpen
  \bibfield  {author} {\bibinfo {author} {\bibfnamefont {F.}~\bibnamefont {Gelis}}, \bibinfo {author} {\bibfnamefont {E.}~\bibnamefont {Iancu}}, \bibinfo {author} {\bibfnamefont {J.}~\bibnamefont {Jalilian-Marian}},\ and\ \bibinfo {author} {\bibfnamefont {R.}~\bibnamefont {Venugopalan}},\ }\bibfield  {title} {\bibinfo {title} {{The Color Glass Condensate}},\ }\href {https://doi.org/10.1146/annurev.nucl.010909.083629} {\bibfield  {journal} {\bibinfo  {journal} {Ann. Rev. Nucl. Part. Sci.}\ }\textbf {\bibinfo {volume} {60}},\ \bibinfo {pages} {463} (\bibinfo {year} {2010})},\ \Eprint {https://arxiv.org/abs/1002.0333} {arXiv:1002.0333 [hep-ph]} \BibitemShut {NoStop}%
\bibitem [{\citenamefont {Iancu}\ and\ \citenamefont {Venugopalan}(2003)}]{Iancu:2003xm}%
  \BibitemOpen
  \bibfield  {author} {\bibinfo {author} {\bibfnamefont {E.}~\bibnamefont {Iancu}}\ and\ \bibinfo {author} {\bibfnamefont {R.}~\bibnamefont {Venugopalan}},\ }\bibinfo {title} {{The Color glass condensate and high-energy scattering in QCD}},\ in\ \href {https://doi.org/10.1142/9789812795533_0005} {\emph {\bibinfo {booktitle} {{Quark-gluon plasma 4}}}},\ \bibinfo {editor} {edited by\ \bibinfo {editor} {\bibfnamefont {R.~C.}\ \bibnamefont {Hwa}}\ and\ \bibinfo {editor} {\bibfnamefont {X.-N.}\ \bibnamefont {Wang}}}\ (\bibinfo {year} {2003})\ pp.\ \bibinfo {pages} {249--3363},\ \Eprint {https://arxiv.org/abs/hep-ph/0303204} {arXiv:hep-ph/0303204} \BibitemShut {NoStop}%
\bibitem [{\citenamefont {McLerran}(2009)}]{McLerran:2008es}%
  \BibitemOpen
  \bibfield  {author} {\bibinfo {author} {\bibfnamefont {L.}~\bibnamefont {McLerran}},\ }\bibfield  {title} {\bibinfo {title} {{A Brief Introduction to the Color Glass Condensate and the Glasma}},\ }in\ \href {https://doi.org/10.3204/DESY-PROC-2009-01/26} {\emph {\bibinfo {booktitle} {{38th International Symposium on Multiparticle Dynamics}}}}\ (\bibinfo {year} {2009})\ pp.\ \bibinfo {pages} {3--18},\ \Eprint {https://arxiv.org/abs/0812.4989} {arXiv:0812.4989 [hep-ph]} \BibitemShut {NoStop}%
\bibitem [{\citenamefont {Gelis}(2013)}]{Gelis:2012ri}%
  \BibitemOpen
  \bibfield  {author} {\bibinfo {author} {\bibfnamefont {F.}~\bibnamefont {Gelis}},\ }\bibfield  {title} {\bibinfo {title} {{Color Glass Condensate and Glasma}},\ }\href {https://doi.org/10.1142/S0217751X13300019} {\bibfield  {journal} {\bibinfo  {journal} {Int. J. Mod. Phys. A}\ }\textbf {\bibinfo {volume} {28}},\ \bibinfo {pages} {1330001} (\bibinfo {year} {2013})},\ \Eprint {https://arxiv.org/abs/1211.3327} {arXiv:1211.3327 [hep-ph]} \BibitemShut {NoStop}%
\bibitem [{\citenamefont {Avramescu}\ \emph {et~al.}(2025)\citenamefont {Avramescu}, \citenamefont {Greco}, \citenamefont {Lappi}, \citenamefont {M{\"a}ntysaari},\ and\ \citenamefont {M{\"u}ller}}]{Avramescu:2024xts}%
  \BibitemOpen
  \bibfield  {author} {\bibinfo {author} {\bibfnamefont {D.}~\bibnamefont {Avramescu}}, \bibinfo {author} {\bibfnamefont {V.}~\bibnamefont {Greco}}, \bibinfo {author} {\bibfnamefont {T.}~\bibnamefont {Lappi}}, \bibinfo {author} {\bibfnamefont {H.}~\bibnamefont {M{\"a}ntysaari}},\ and\ \bibinfo {author} {\bibfnamefont {D.}~\bibnamefont {M{\"u}ller}},\ }\bibfield  {title} {\bibinfo {title} {{Heavy-Flavor Angular Correlations as a Direct Probe of the Glasma}},\ }\href {https://doi.org/10.1103/PhysRevLett.134.172301} {\bibfield  {journal} {\bibinfo  {journal} {Phys. Rev. Lett.}\ }\textbf {\bibinfo {volume} {134}},\ \bibinfo {pages} {172301} (\bibinfo {year} {2025})},\ \Eprint {https://arxiv.org/abs/2409.10565} {arXiv:2409.10565 [hep-ph]} \BibitemShut {NoStop}%
\bibitem [{\citenamefont {Oliva}\ \emph {et~al.}(2025)\citenamefont {Oliva}, \citenamefont {Parisi}, \citenamefont {Greco},\ and\ \citenamefont {Ruggieri}}]{Oliva:2024rex}%
  \BibitemOpen
  \bibfield  {author} {\bibinfo {author} {\bibfnamefont {L.}~\bibnamefont {Oliva}}, \bibinfo {author} {\bibfnamefont {G.}~\bibnamefont {Parisi}}, \bibinfo {author} {\bibfnamefont {V.}~\bibnamefont {Greco}},\ and\ \bibinfo {author} {\bibfnamefont {M.}~\bibnamefont {Ruggieri}},\ }\bibfield  {title} {\bibinfo {title} {{Melting of cc{\textasciimacron} and bb{\textasciimacron} pairs in the pre-equilibrium stage of proton-nucleus collisions at the Large Hadron Collider}},\ }\href {https://doi.org/10.1103/nc3z-vns9} {\bibfield  {journal} {\bibinfo  {journal} {Phys. Rev. D}\ }\textbf {\bibinfo {volume} {112}},\ \bibinfo {pages} {014008} (\bibinfo {year} {2025})},\ \Eprint {https://arxiv.org/abs/2412.07967} {arXiv:2412.07967 [hep-ph]} \BibitemShut {NoStop}%
\bibitem [{\citenamefont {Pooja}\ \emph {et~al.}(2023)\citenamefont {Pooja}, \citenamefont {Das}, \citenamefont {Greco},\ and\ \citenamefont {Ruggieri}}]{Pooja:2023gqt}%
  \BibitemOpen
  \bibfield  {author} {\bibinfo {author} {\bibnamefont {Pooja}}, \bibinfo {author} {\bibfnamefont {S.~K.}\ \bibnamefont {Das}}, \bibinfo {author} {\bibfnamefont {V.}~\bibnamefont {Greco}},\ and\ \bibinfo {author} {\bibfnamefont {M.}~\bibnamefont {Ruggieri}},\ }\bibfield  {title} {\bibinfo {title} {{Thermalization and isotropization of heavy quarks in a non-Markovian medium in high-energy nuclear collisions}},\ }\href {https://doi.org/10.1103/PhysRevD.108.054026} {\bibfield  {journal} {\bibinfo  {journal} {Phys. Rev. D}\ }\textbf {\bibinfo {volume} {108}},\ \bibinfo {pages} {054026} (\bibinfo {year} {2023})},\ \Eprint {https://arxiv.org/abs/2306.13749} {arXiv:2306.13749 [hep-ph]} \BibitemShut {NoStop}%
\bibitem [{\citenamefont {Avramescu}\ \emph {et~al.}(2023)\citenamefont {Avramescu}, \citenamefont {B{\u{a}}ran}, \citenamefont {Greco}, \citenamefont {Ipp}, \citenamefont {M{\"u}ller},\ and\ \citenamefont {Ruggieri}}]{Avramescu:2023qvv}%
  \BibitemOpen
  \bibfield  {author} {\bibinfo {author} {\bibfnamefont {D.}~\bibnamefont {Avramescu}}, \bibinfo {author} {\bibfnamefont {V.}~\bibnamefont {B{\u{a}}ran}}, \bibinfo {author} {\bibfnamefont {V.}~\bibnamefont {Greco}}, \bibinfo {author} {\bibfnamefont {A.}~\bibnamefont {Ipp}}, \bibinfo {author} {\bibfnamefont {D.~I.}\ \bibnamefont {M{\"u}ller}},\ and\ \bibinfo {author} {\bibfnamefont {M.}~\bibnamefont {Ruggieri}},\ }\bibfield  {title} {\bibinfo {title} {{Simulating jets and heavy quarks in the glasma using the colored particle-in-cell method}},\ }\href {https://doi.org/10.1103/PhysRevD.107.114021} {\bibfield  {journal} {\bibinfo  {journal} {Phys. Rev. D}\ }\textbf {\bibinfo {volume} {107}},\ \bibinfo {pages} {114021} (\bibinfo {year} {2023})},\ \Eprint {https://arxiv.org/abs/2303.05599} {arXiv:2303.05599 [hep-ph]} \BibitemShut {NoStop}%
\bibitem [{\citenamefont {Sun}\ \emph {et~al.}(2019)\citenamefont {Sun}, \citenamefont {Coci}, \citenamefont {Das}, \citenamefont {Plumari}, \citenamefont {Ruggieri},\ and\ \citenamefont {Greco}}]{Sun:2019fud}%
  \BibitemOpen
  \bibfield  {author} {\bibinfo {author} {\bibfnamefont {Y.}~\bibnamefont {Sun}}, \bibinfo {author} {\bibfnamefont {G.}~\bibnamefont {Coci}}, \bibinfo {author} {\bibfnamefont {S.~K.}\ \bibnamefont {Das}}, \bibinfo {author} {\bibfnamefont {S.}~\bibnamefont {Plumari}}, \bibinfo {author} {\bibfnamefont {M.}~\bibnamefont {Ruggieri}},\ and\ \bibinfo {author} {\bibfnamefont {V.}~\bibnamefont {Greco}},\ }\bibfield  {title} {\bibinfo {title} {{Impact of Glasma on heavy quark observables in nucleus-nucleus collisions at LHC}},\ }\href {https://doi.org/10.1016/j.physletb.2019.134933} {\bibfield  {journal} {\bibinfo  {journal} {Phys. Lett. B}\ }\textbf {\bibinfo {volume} {798}},\ \bibinfo {pages} {134933} (\bibinfo {year} {2019})},\ \Eprint {https://arxiv.org/abs/1902.06254} {arXiv:1902.06254 [nucl-th]} \BibitemShut {NoStop}%
\bibitem [{\citenamefont {Liu}\ \emph {et~al.}(2020)\citenamefont {Liu}, \citenamefont {Plumari}, \citenamefont {Das}, \citenamefont {Greco},\ and\ \citenamefont {Ruggieri}}]{Liu:2019lac}%
  \BibitemOpen
  \bibfield  {author} {\bibinfo {author} {\bibfnamefont {J.~H.}\ \bibnamefont {Liu}}, \bibinfo {author} {\bibfnamefont {S.}~\bibnamefont {Plumari}}, \bibinfo {author} {\bibfnamefont {S.~K.}\ \bibnamefont {Das}}, \bibinfo {author} {\bibfnamefont {V.}~\bibnamefont {Greco}},\ and\ \bibinfo {author} {\bibfnamefont {M.}~\bibnamefont {Ruggieri}},\ }\bibfield  {title} {\bibinfo {title} {{Diffusion of heavy quarks in the early stage of high-energy nuclear collisions at energies available at the BNL Relativistic Heavy Ion Collider and at the CERN Large Hadron Collider}},\ }\href {https://doi.org/10.1103/PhysRevC.102.044902} {\bibfield  {journal} {\bibinfo  {journal} {Phys. Rev. C}\ }\textbf {\bibinfo {volume} {102}},\ \bibinfo {pages} {044902} (\bibinfo {year} {2020})},\ \Eprint {https://arxiv.org/abs/1911.02480} {arXiv:1911.02480 [nucl-th]} \BibitemShut {NoStop}%
\bibitem [{\citenamefont {Boguslavski}\ \emph {et~al.}(2020)\citenamefont {Boguslavski}, \citenamefont {Kurkela}, \citenamefont {Lappi},\ and\ \citenamefont {Peuron}}]{Boguslavski:2020tqz}%
  \BibitemOpen
  \bibfield  {author} {\bibinfo {author} {\bibfnamefont {K.}~\bibnamefont {Boguslavski}}, \bibinfo {author} {\bibfnamefont {A.}~\bibnamefont {Kurkela}}, \bibinfo {author} {\bibfnamefont {T.}~\bibnamefont {Lappi}},\ and\ \bibinfo {author} {\bibfnamefont {J.}~\bibnamefont {Peuron}},\ }\bibfield  {title} {\bibinfo {title} {{Heavy quark diffusion in an overoccupied gluon plasma}},\ }\href {https://doi.org/10.1007/JHEP09(2020)077} {\bibfield  {journal} {\bibinfo  {journal} {JHEP}\ }\textbf {\bibinfo {volume} {09}},\ \bibinfo {pages} {077}},\ \Eprint {https://arxiv.org/abs/2005.02418} {arXiv:2005.02418 [hep-ph]} \BibitemShut {NoStop}%
\bibitem [{\citenamefont {Carrington}\ \emph {et~al.}(2022{\natexlab{a}})\citenamefont {Carrington}, \citenamefont {Czajka},\ and\ \citenamefont {Mrowczynski}}]{Carrington:2020ssh}%
  \BibitemOpen
  \bibfield  {author} {\bibinfo {author} {\bibfnamefont {M.~E.}\ \bibnamefont {Carrington}}, \bibinfo {author} {\bibfnamefont {A.}~\bibnamefont {Czajka}},\ and\ \bibinfo {author} {\bibfnamefont {S.}~\bibnamefont {Mrowczynski}},\ }\bibfield  {title} {\bibinfo {title} {{The energy-momentum tensor at the earliest stage of relativistic heavy-ion collisions}},\ }\href {https://doi.org/10.1140/epja/s10050-021-00600-x} {\bibfield  {journal} {\bibinfo  {journal} {Eur. Phys. J. A}\ }\textbf {\bibinfo {volume} {58}},\ \bibinfo {pages} {5} (\bibinfo {year} {2022}{\natexlab{a}})},\ \Eprint {https://arxiv.org/abs/2012.03042} {arXiv:2012.03042 [hep-ph]} \BibitemShut {NoStop}%
\bibitem [{\citenamefont {Carrington}\ \emph {et~al.}(2022{\natexlab{b}})\citenamefont {Carrington}, \citenamefont {Czajka},\ and\ \citenamefont {Mr{\'o}wczy{\'n}ski}}]{Carrington:2021qvi}%
  \BibitemOpen
  \bibfield  {author} {\bibinfo {author} {\bibfnamefont {M.~E.}\ \bibnamefont {Carrington}}, \bibinfo {author} {\bibfnamefont {A.}~\bibnamefont {Czajka}},\ and\ \bibinfo {author} {\bibfnamefont {S.}~\bibnamefont {Mr{\'o}wczy{\'n}ski}},\ }\bibfield  {title} {\bibinfo {title} {{Physical characteristics of glasma from the earliest stage of relativistic heavy ion collisions}},\ }\href {https://doi.org/10.1103/PhysRevC.106.034904} {\bibfield  {journal} {\bibinfo  {journal} {Phys. Rev. C}\ }\textbf {\bibinfo {volume} {106}},\ \bibinfo {pages} {034904} (\bibinfo {year} {2022}{\natexlab{b}})},\ \Eprint {https://arxiv.org/abs/2105.05327} {arXiv:2105.05327 [hep-ph]} \BibitemShut {NoStop}%
\bibitem [{\citenamefont {Iida}\ \emph {et~al.}(2014)\citenamefont {Iida}, \citenamefont {Kunihiro}, \citenamefont {Ohnishi},\ and\ \citenamefont {Takahashi}}]{Iida:2014wea}%
  \BibitemOpen
  \bibfield  {author} {\bibinfo {author} {\bibfnamefont {H.}~\bibnamefont {Iida}}, \bibinfo {author} {\bibfnamefont {T.}~\bibnamefont {Kunihiro}}, \bibinfo {author} {\bibfnamefont {A.}~\bibnamefont {Ohnishi}},\ and\ \bibinfo {author} {\bibfnamefont {T.~T.}\ \bibnamefont {Takahashi}},\ }\bibfield  {title} {\bibinfo {title} {{Time evolution of gluon coherent state and its von Neumann entropy in heavy-ion collisions}},\ }\href@noop {} {\bibfield  {journal} {\bibinfo  {journal} {{}}\ } (\bibinfo {year} {2014})},\ \Eprint {https://arxiv.org/abs/1410.7309} {arXiv:1410.7309 [hep-ph]} \BibitemShut {NoStop}%
\bibitem [{\citenamefont {Matsuda}\ \emph {et~al.}(2022)\citenamefont {Matsuda}, \citenamefont {Kunihiro}, \citenamefont {Ohnishi},\ and\ \citenamefont {Takahashi}}]{Matsuda:2022hok}%
  \BibitemOpen
  \bibfield  {author} {\bibinfo {author} {\bibfnamefont {H.}~\bibnamefont {Matsuda}}, \bibinfo {author} {\bibfnamefont {T.}~\bibnamefont {Kunihiro}}, \bibinfo {author} {\bibfnamefont {A.}~\bibnamefont {Ohnishi}},\ and\ \bibinfo {author} {\bibfnamefont {T.~T.}\ \bibnamefont {Takahashi}},\ }\bibfield  {title} {\bibinfo {title} {{Entropy production in a longitudinally expanding Yang{\textendash}Mills field with use of the Husimi function: semiclassical approximation}},\ }\href {https://doi.org/10.1093/ptep/ptac086} {\bibfield  {journal} {\bibinfo  {journal} {PTEP}\ }\textbf {\bibinfo {volume} {2022}},\ \bibinfo {pages} {073D02} (\bibinfo {year} {2022})},\ \Eprint {https://arxiv.org/abs/2203.02859} {arXiv:2203.02859 [hep-ph]} \BibitemShut {NoStop}%
\bibitem [{\citenamefont {Tsukiji}\ \emph {et~al.}(2018)\citenamefont {Tsukiji}, \citenamefont {Kunihiro}, \citenamefont {Ohnishi},\ and\ \citenamefont {Takahashi}}]{Tsukiji:2017pjx}%
  \BibitemOpen
  \bibfield  {author} {\bibinfo {author} {\bibfnamefont {H.}~\bibnamefont {Tsukiji}}, \bibinfo {author} {\bibfnamefont {T.}~\bibnamefont {Kunihiro}}, \bibinfo {author} {\bibfnamefont {A.}~\bibnamefont {Ohnishi}},\ and\ \bibinfo {author} {\bibfnamefont {T.~T.}\ \bibnamefont {Takahashi}},\ }\bibfield  {title} {\bibinfo {title} {{Entropy production and isotropization in Yang{\textendash}Mills theory using a quantum distribution function}},\ }\href {https://doi.org/10.1093/ptep/ptx186} {\bibfield  {journal} {\bibinfo  {journal} {PTEP}\ }\textbf {\bibinfo {volume} {2018}},\ \bibinfo {pages} {013D02} (\bibinfo {year} {2018})},\ \Eprint {https://arxiv.org/abs/1709.00979} {arXiv:1709.00979 [hep-ph]} \BibitemShut {NoStop}%
\bibitem [{\citenamefont {Tsukiji}\ \emph {et~al.}(2016)\citenamefont {Tsukiji}, \citenamefont {Iida}, \citenamefont {Kunihiro}, \citenamefont {Ohnishi},\ and\ \citenamefont {Takahashi}}]{Tsukiji:2016krj}%
  \BibitemOpen
  \bibfield  {author} {\bibinfo {author} {\bibfnamefont {H.}~\bibnamefont {Tsukiji}}, \bibinfo {author} {\bibfnamefont {H.}~\bibnamefont {Iida}}, \bibinfo {author} {\bibfnamefont {T.}~\bibnamefont {Kunihiro}}, \bibinfo {author} {\bibfnamefont {A.}~\bibnamefont {Ohnishi}},\ and\ \bibinfo {author} {\bibfnamefont {T.~T.}\ \bibnamefont {Takahashi}},\ }\bibfield  {title} {\bibinfo {title} {{Entropy production from chaoticity in Yang-Mills field theory with use of the Husimi function}},\ }\href {https://doi.org/10.1103/PhysRevD.94.091502} {\bibfield  {journal} {\bibinfo  {journal} {Phys. Rev. D}\ }\textbf {\bibinfo {volume} {94}},\ \bibinfo {pages} {091502} (\bibinfo {year} {2016})},\ \Eprint {https://arxiv.org/abs/1603.04622} {arXiv:1603.04622 [hep-ph]} \BibitemShut {NoStop}%
\bibitem [{\citenamefont {Iida}\ \emph {et~al.}(2013)\citenamefont {Iida}, \citenamefont {Kunihiro}, \citenamefont {Mueller}, \citenamefont {Ohnishi}, \citenamefont {Schaefer},\ and\ \citenamefont {Takahashi}}]{Iida:2013qwa}%
  \BibitemOpen
  \bibfield  {author} {\bibinfo {author} {\bibfnamefont {H.}~\bibnamefont {Iida}}, \bibinfo {author} {\bibfnamefont {T.}~\bibnamefont {Kunihiro}}, \bibinfo {author} {\bibfnamefont {B.}~\bibnamefont {Mueller}}, \bibinfo {author} {\bibfnamefont {A.}~\bibnamefont {Ohnishi}}, \bibinfo {author} {\bibfnamefont {A.}~\bibnamefont {Schaefer}},\ and\ \bibinfo {author} {\bibfnamefont {T.~T.}\ \bibnamefont {Takahashi}},\ }\bibfield  {title} {\bibinfo {title} {{Entropy production in classical Yang-Mills theory from Glasma initial conditions}},\ }\href {https://doi.org/10.1103/PhysRevD.88.094006} {\bibfield  {journal} {\bibinfo  {journal} {Phys. Rev. D}\ }\textbf {\bibinfo {volume} {88}},\ \bibinfo {pages} {094006} (\bibinfo {year} {2013})},\ \Eprint {https://arxiv.org/abs/1304.1807} {arXiv:1304.1807 [hep-ph]} \BibitemShut {NoStop}%
\bibitem [{\citenamefont {Muller}\ and\ \citenamefont {Schafer}(2006)}]{Muller:2005yu}%
  \BibitemOpen
  \bibfield  {author} {\bibinfo {author} {\bibfnamefont {B.}~\bibnamefont {Muller}}\ and\ \bibinfo {author} {\bibfnamefont {A.}~\bibnamefont {Schafer}},\ }\bibfield  {title} {\bibinfo {title} {{The Decoherence time in high energy heavy ion collisions}},\ }\href {https://doi.org/10.1103/PhysRevC.73.054905} {\bibfield  {journal} {\bibinfo  {journal} {Phys. Rev. C}\ }\textbf {\bibinfo {volume} {73}},\ \bibinfo {pages} {054905} (\bibinfo {year} {2006})},\ \Eprint {https://arxiv.org/abs/hep-ph/0512100} {arXiv:hep-ph/0512100} \BibitemShut {NoStop}%
\bibitem [{\citenamefont {Glauber}(1963)}]{Glauber:1963fi}%
  \BibitemOpen
  \bibfield  {author} {\bibinfo {author} {\bibfnamefont {R.~J.}\ \bibnamefont {Glauber}},\ }\bibfield  {title} {\bibinfo {title} {{The Quantum theory of optical coherence}},\ }\href {https://doi.org/10.1103/PhysRev.130.2529} {\bibfield  {journal} {\bibinfo  {journal} {Phys. Rev.}\ }\textbf {\bibinfo {volume} {130}},\ \bibinfo {pages} {2529} (\bibinfo {year} {1963})}\BibitemShut {NoStop}%
\bibitem [{\citenamefont {Walls}\ and\ \citenamefont {Milburn}(1985)}]{Walls:1985tm}%
  \BibitemOpen
  \bibfield  {author} {\bibinfo {author} {\bibfnamefont {D.~F.}\ \bibnamefont {Walls}}\ and\ \bibinfo {author} {\bibfnamefont {G.~J.}\ \bibnamefont {Milburn}},\ }\bibfield  {title} {\bibinfo {title} {Effect of dissipation on quantum coherence},\ }\href {https://doi.org/10.1103/PhysRevA.31.2403} {\bibfield  {journal} {\bibinfo  {journal} {Phys. Rev. A}\ }\textbf {\bibinfo {volume} {31}},\ \bibinfo {pages} {2403} (\bibinfo {year} {1985})}\BibitemShut {NoStop}%
\bibitem [{\citenamefont {Schenke}\ \emph {et~al.}(2020)\citenamefont {Schenke}, \citenamefont {Shen},\ and\ \citenamefont {Tribedy}}]{Schenke:2020mbo}%
  \BibitemOpen
  \bibfield  {author} {\bibinfo {author} {\bibfnamefont {B.}~\bibnamefont {Schenke}}, \bibinfo {author} {\bibfnamefont {C.}~\bibnamefont {Shen}},\ and\ \bibinfo {author} {\bibfnamefont {P.}~\bibnamefont {Tribedy}},\ }\bibfield  {title} {\bibinfo {title} {{Running the gamut of high energy nuclear collisions}},\ }\href {https://doi.org/10.1103/PhysRevC.102.044905} {\bibfield  {journal} {\bibinfo  {journal} {Phys. Rev. C}\ }\textbf {\bibinfo {volume} {102}},\ \bibinfo {pages} {044905} (\bibinfo {year} {2020})},\ \Eprint {https://arxiv.org/abs/2005.14682} {arXiv:2005.14682 [nucl-th]} \BibitemShut {NoStop}%
\bibitem [{\citenamefont {Parisi}\ \emph {et~al.}(2025)\citenamefont {Parisi}, \citenamefont {Greco},\ and\ \citenamefont {Ruggieri}}]{Parisi:2025slf}%
  \BibitemOpen
  \bibfield  {author} {\bibinfo {author} {\bibfnamefont {G.}~\bibnamefont {Parisi}}, \bibinfo {author} {\bibfnamefont {V.}~\bibnamefont {Greco}},\ and\ \bibinfo {author} {\bibfnamefont {M.}~\bibnamefont {Ruggieri}},\ }\bibfield  {title} {\bibinfo {title} {{Anisotropic fluctuations of momentum and angular momentum of heavy quarks in the pre-equilibrium stage of pA collisions at the LHC}},\ }\href@noop {} {\bibfield  {journal} {\bibinfo  {journal} {{}}\ } (\bibinfo {year} {2025})},\ \Eprint {https://arxiv.org/abs/2505.08441} {arXiv:2505.08441 [hep-ph]} \BibitemShut {NoStop}%
\bibitem [{\citenamefont {Rezaeian}\ \emph {et~al.}(2013)\citenamefont {Rezaeian}, \citenamefont {Siddikov}, \citenamefont {Van~de Klundert},\ and\ \citenamefont {Venugopalan}}]{Rezaeian:2012ji}%
  \BibitemOpen
  \bibfield  {author} {\bibinfo {author} {\bibfnamefont {A.~H.}\ \bibnamefont {Rezaeian}}, \bibinfo {author} {\bibfnamefont {M.}~\bibnamefont {Siddikov}}, \bibinfo {author} {\bibfnamefont {M.}~\bibnamefont {Van~de Klundert}},\ and\ \bibinfo {author} {\bibfnamefont {R.}~\bibnamefont {Venugopalan}},\ }\bibfield  {title} {\bibinfo {title} {{Analysis of combined HERA data in the Impact-Parameter dependent Saturation model}},\ }\href {https://doi.org/10.1103/PhysRevD.87.034002} {\bibfield  {journal} {\bibinfo  {journal} {Phys. Rev. D}\ }\textbf {\bibinfo {volume} {87}},\ \bibinfo {pages} {034002} (\bibinfo {year} {2013})},\ \Eprint {https://arxiv.org/abs/1212.2974} {arXiv:1212.2974 [hep-ph]} \BibitemShut {NoStop}%
\bibitem [{\citenamefont {Lappi}(2008)}]{Lappi:2007ku}%
  \BibitemOpen
  \bibfield  {author} {\bibinfo {author} {\bibfnamefont {T.}~\bibnamefont {Lappi}},\ }\bibfield  {title} {\bibinfo {title} {{Wilson line correlator in the MV model: Relating the glasma to deep inelastic scattering}},\ }\href {https://doi.org/10.1140/epjc/s10052-008-0588-4} {\bibfield  {journal} {\bibinfo  {journal} {Eur. Phys. J. C}\ }\textbf {\bibinfo {volume} {55}},\ \bibinfo {pages} {285} (\bibinfo {year} {2008})},\ \Eprint {https://arxiv.org/abs/0711.3039} {arXiv:0711.3039 [hep-ph]} \BibitemShut {NoStop}%
\end{thebibliography}%

\end{document}